\documentclass[%
 amsmath,amssymb,
 aps,
floatfix,
]{revtex4-2}

\usepackage{graphicx}
\usepackage{dcolumn}
\usepackage{bm}
\usepackage{array}%
\usepackage{listings}
\usepackage{xspace}
\usepackage{subcaption}
\usepackage{booktabs}
\usepackage{multirow}
\usepackage{makecell}
\usepackage{siunitx}
\usepackage{hyperref}

\newcommand{\herwig}{{\tt Herwig 7}\xspace}
\newcommand{\Matchbox}{{\tt Matchbox}\xspace}
\newcommand{\vbfnlo}{{\tt VBFNLO 3}\xspace}
\newcommand{\GeV}{\ensuremath{\,\mathrm{GeV}}\xspace}
\newcommand{\TeV}{\ensuremath{\,\mathrm{TeV}}\xspace}

\newcommand{\as}{\ensuremath{\alpha_s}}

\newcommand{\hjjlops}{\ensuremath{h(2)\bigoplus {\rm PS}}~}
\newcommand{\hjjnlops}{\ensuremath{h(2^\star)\bigoplus {\rm PS}}~}
\newcommand{\hjjnlo}{\ensuremath{h(2^\star)}~}
\newcommand{\hjjmerge}[1][3^\star,4]{\ensuremath{h(2^\star, #1)}~}

\newcommand{\dphijj}{\ensuremath{|\Delta\phi_{j_1j_2}|}~}
\newcommand{\dphijfjb}{\ensuremath{\Delta\phi_{j_fj_b}}~}
\newcommand{\mjj}{\ensuremath{m_{j_1j_2}}~}
\newcommand{\dyjj}{\ensuremath{\Delta y_{j_1j_2}}~}

\begin{document}


\title{Next-to-Leading-Order Multi-Jet Merging of Higgs Boson Production via Vector Boson Fusion in the Presence of Anomalous Couplings}

\author{Tinghua Chen}
\affiliation{ IT Research Cyberinfrastructure,
 University of Delaware\\
  Newark, DE 19716, USA\\
}%

\author{Terrance Figy}
\affiliation{%
 Department of Mathematics,Statistics, and Physics, Wichita State University  \\
 Wichita, Kansas 67260, USA \\
}%

\author{Tunde Kushimo}
\affiliation{%
 Department of Mathematics,Statistics, and Physics, Wichita State University \\
 Wichita, Kansas 67260, USA \\
}%

\date{\today}


\begin{abstract}
We present the results of a detailed simulation study of Higgs boson production via vector boson fusion (VBF) at the Large Hadron Collider (LHC) in the presence of anomalous couplings. The analysis utilizes the \texttt{Herwig} event generator interfaced with \texttt{VBFNLO}, employing a multi-jet merging scheme that combines NLO accurate matrix elements for Higgs plus 2- and 3-jet multiplicities with LO accurate matrix elements for Higgs plus 4-jet multiplicities. We investigate both matching and merging setups to consistently combine these hard scattering processes with QCD parton showers. Finally, we demonstrate the robustness of azimuthal angle observables against QCD radiative corrections across these frameworks.
\end{abstract}

\maketitle


\section{\label{sec:intro} Introduction}

The discovery of the Higgs boson at the Large Hadron Collider (LHC) in 2012 confirmed the mechanism of electroweak symmetry breaking predicted by the Standard Model (SM) and marked a major milestone in particle physics~\cite{ATLAS:2012yve, CMS:2012qbp}. Since then, a central objective of the LHC physics program has been the precise determination of the properties of the Higgs boson, in particular its couplings to other fundamental particles~\cite{ATLAS:2022vkf}. Among these, the interaction between the Higgs boson and the electroweak gauge bosons $W$ and $Z$, commonly referred to as the $HVV$ coupling, plays a crucial role~\cite{Hagiwara:1993qt, Hankele:2006ma}. In the SM, this coupling is CP-even and fully fixed by the gauge structure of the theory. However, extensions of the SM allow for deviations in both its strength and CP structure, potentially introducing CP-odd or CP-mixed contributions that would constitute clear signals of physics beyond the SM~\cite{Hagiwara:1993ck, Contino:2013kra, Passarino:2013nka}.

The study of CP-sensitive observables is therefore of fundamental importance in Higgs physics, as they provide direct access to the CP nature of the Higgs boson’s interactions. Unlike inclusive production rates, which are often only weakly affected by CP-violating effects, differential observables such as angular correlations and azimuthal separations of final-state particles can exhibit characteristic distortions in the presence of CP-odd contributions~\cite{Plehn:2001nj}. These observables offer a powerful and largely model-independent means of probing CP violation in the Higgs sector, complementing total cross-section measurements and allowing sensitivity to interference effects between CP-even and CP-odd amplitudes~\cite{Englert:2014uua, Contino:2013kra}.

Vector Boson Fusion (VBF) is one of the most powerful Higgs production mechanisms to probe the $HVV$ interaction in the LHC. In this process, two incoming quarks radiate weak gauge bosons that subsequently fuse into a Higgs boson, resulting in a distinctive final state characterized by two energetic forward tagging jets with a large rapidity separation and suppressed hadronic activity in the central region. These kinematic features render VBF particularly well suited for precision studies of Higgs couplings~\cite{Barone:2025jey}. Early theoretical investigations demonstrated that angular correlations between the tagging jets, most notably the azimuthal angle separation, are highly sensitive to the CP structure of the Higgs interaction~\cite{Figy:2004pt, Plehn:2001nj}. Such observables provide a robust handle for discriminating among CP-even, CP-odd, and CP-mixed scenarios.

The theoretical description of VBF Higgs boson production has undergone continuous refinement over the past two decades. Early next-to-leading-order (NLO) QCD calculations established that QCD corrections to the electroweak VBF process are generally modest, leading to corrections of only a few percent for inclusive cross sections while substantially reducing renormalization and factorization-scale uncertainties \cite{Han:1992hr,Figy:2003nv, Figy:2004pt, Berger:2004pca}. These studies also demonstrated that the characteristic VBF topology, consisting of two widely separated forward tagging jets with little color exchange between them, remains remarkably stable under higher-order QCD corrections. Subsequent developments incorporated realistic Higgs decays, differential distributions, and flexible Monte Carlo implementations within parton-level generators such as \vbfnlo, providing the precision tools required for detailed phenomenological studies and direct comparisons with LHC measurements \cite{Arnold:2008rz, Baglio:2014uba,Baglio:2024gyp}. More recently, precision studies have extended beyond fixed-order calculations to investigate the impact of jet clustering, parton-shower matching, and realistic event simulation on VBF observables, further improving the theoretical description of Higgs production in experimentally relevant environments \cite{Rauch:2017JetClustering,Rauch:2017VBFWW}. The next-to-next-to-leading order (NNLO) QCD corrections for VBF Higgs boson production for total cross-sections were determined using a structure function approach \cite{Bolzoni:2010xr} and later predictions for differential cross-sections become available \cite{Cacciari:2015jma,Cruz-Martinez:2018rod}. The next-to-next-to-next-to leading order QCD correctons are also avaialble \cite{Dreyer:2016oyx}.

To exploit this sensitivity reliably, accurate theoretical predictions for VBF Higgs production are essential, requiring the inclusion of higher-order QCD effects. NLO QCD corrections to VBF processes are known to stabilize total cross sections, reduce scale uncertainties, and induce non-negligible effects in differential distributions relevant for experimental analyses~\cite{Figy:2003nv,Berger:2004pca}. In realistic collider environments, these fixed-order calculations must be combined with parton showers to account for multiple soft and collinear emissions. Modern Monte Carlo event generators provide such capabilities, with tools like \vbfnlo~\cite{Arnold:2008rz} supplying NLO matrix elements for VBF processes and general-purpose generators such as \herwig~\cite{Bahr:2008pv} enabling consistent NLO matching and multi-jet merging, enabling realistic simulations of additional QCD radiation \cite{platzer2012dipole,Bellm:2017ktr}.

The full NLO QCD and electroweak corrections to electroweak Higgs boson production in association with two jets were computed in Ref.~\cite{Ciccolini:2007jr}. Several groups have also presented NLO+parton-shower matched predictions for electroweak Higgs boson production with two jets \cite{DErrico:2011wfa,Frixione:2013mta,Nason:2009ai,Jager:2020hkz}. The NLO QCD corrections to VBF Higgs boson production in association with three jets were first presented in Ref.~\cite{Figy:2007kv}. The QCD corrections were subsequently computed for the electroweak Higgs boson production in association of three-jets where all Feynman diagram topologies were included in Refs.~\cite{Campanario:2013nca,Campanario:2013fsa,Campanario:2014aia,Campanario:2018ppz}. NLO+parton-shower matched predictions for electroweak Higgs boson production with three jets were presented in Refs.~\cite{Jager:2014vna,Azzi:2019yne}, while both matched and merged predictions for the full electroweak and VBF approximations at NLO were presented in Ref.~\cite{Chen:2021phj}. Reference~\cite{Hoche:2021mkv} discusses several parton-shower algorithms in the context of {\tt Pythia}~\cite{Sjostrand:2014zea} and {\tt Vincia}~\cite{Brooks:2020upa}. For a review of the current state of the art in VBF Higgs boson production, see Ref.~\cite{Barone:2025jey}. 

NLO matching schemes provide a consistent combination of fixed-order calculations and parton showers while preserving NLO accuracy for inclusive observables. Multi-jet merging techniques further extend this framework by incorporating matrix elements with different jet multiplicities, thereby improving the description of hard additional radiation. The theoretical foundations and practical implementations of these approaches have been extensively studied~\cite{hoche2016introduction, Alwall:2011uj}. While inclusive cross sections are typically only mildly affected, differential observables, particularly those involving jet kinematics and angular correlations, can exhibit significant sensitivity to the treatment of higher-order radiation. This sensitivity becomes especially pronounced in the presence of anomalous $HVV$ couplings, where subtle shape distortions in CP-sensitive observables carry essential information about the underlying interaction structure.

Anomalous Higgs couplings are commonly described using an effective field theory framework, in which higher-dimensional operators supplement the SM Lagrangian. This approach provides a model-independent parametrization of potential new physics effects, including CP-even and CP-odd contributions to the $HVV$ vertex~\cite{Hagiwara:1993ck, Hagiwara:1993qt, Gonzalez-Garcia:1999ije, Contino:2013kra}. Phenomenological studies have demonstrated that these anomalous interactions lead to characteristic modifications of angular and kinematic distributions in VBF events~\cite{Hankele:2006ma, Figy:2004pt, Maltoni:2013sma, Englert:2014uua}. In addition, momentum-dependent form factors are often introduced to regulate the high-energy behavior of these operators and to parametrize the scale at which new physics becomes relevant \cite{Hankele:2006ma, Maltoni:2013sma, Artoisenet:2013puc}.

Despite extensive work on CP-sensitive observables in VBF Higgs production~\cite{Plehn:2001nj, Figy:2003nv, Englert:2014uua, Contino:2013kra, Passarino:2013nka}, a systematic investigation of how anomalous $HVV$ couplings interact with modern NLO-matched and multi-jet merged simulations remains largely unexplored. In particular, the combined impact of CP structure, higher-order QCD radiation, and realistic LHC selection cuts on key observables, such as tagging-jet rapidity gaps and azimuthal correlations, has not been comprehensively quantified. Moreover, the role of parton showers and extra jet emissions in reshaping CP-sensitive distributions, as well as the interplay with form factors, is often neglected or treated inconsistently across existing studies.

The aim of this work is to address these gaps through a detailed simulation study of the VBF Higgs production at the LHC in the presence of anomalous CP-even, CP-odd, and CP-mixed $HVV$ couplings. We study the impact of these anomalous interactions using NLO QCD matrix elements from \vbfnlo, consistently matched and merged with parton showers in \herwig. A central focus is placed on understanding how NLO matching and multi-jet merging algorithms influence jet-related observables and CP-sensitive angular correlations, thereby quantifying the effect of additional QCD radiation on phenomenological predictions. By systematically comparing different simulation setups and analyzing a range of kinematic observables under realistic event selection criteria, we identify robust signatures of CP-violating effects and assess their sensitivity to higher-order QCD corrections, providing valuable guidance for future experimental analyses and precision phenomenology at the LHC.

The remainder of this paper is organized as follows. In Sec.~\ref{sec:calc} and Sec.~\ref{sec:setup}, we describe the details and set-up for our calculations, the computational setup, including the implementation of anomalous $HVV$ couplings, details of the NLO calculations, Monte Carlo input parameters, parton-shower matching and multi-jet merging procedures, and the event selection cuts employed in the analysis. Sec.~\ref{sec:results} presents the phenomenological results, focusing on CP-sensitive observables in VBF Higgs production, systematic comparisons between NLO-matched and NLO-merged predictions, deviations induced by anomalous couplings relative to the SM, and the impact of form factors. Finally, Sec.~\ref{sec:conc} summarizes our findings and outlines their implications for future experimental analyses and precision studies at the LHC.

\section{\label{sec:calc} Details of Calculation and Simulation}
%

\begin{figure}[t]
    \centering
    \setlength{\tabcolsep}{4pt}

    \begin{tabular}{@{}cc@{}}
        \parbox[t][0.37\columnwidth][t]{0.42\columnwidth}{%
            \vspace{0pt}
            \centering
            \includegraphics[width=\linewidth]
                {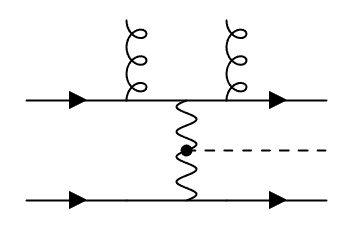}
        }
        &
        \parbox[t][0.37\columnwidth][t]{0.42\columnwidth}{%
            \vspace{0pt}
            \centering
            \includegraphics[width=\linewidth]
                {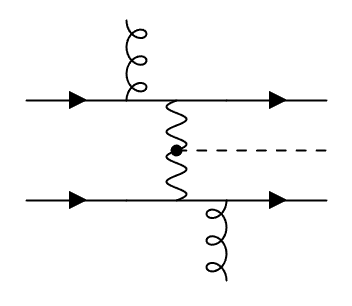}
        }
        \\[-0.4em]
        \textbf{(a)} & \textbf{(b)}
        \\[0.4em]

        \includegraphics[width=0.42\columnwidth]{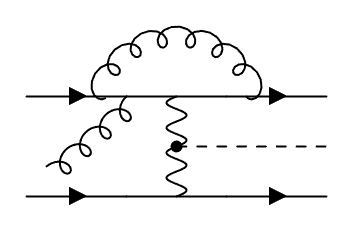}
        &
        \includegraphics[width=0.42\columnwidth]{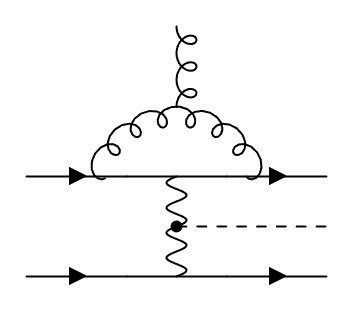}
        \\[-0.4em]
        \textbf{(c)} & \textbf{(d)}
        \\[0.4em]

    \end{tabular}

    \caption{Representative Feynman diagrams for Higgs-boson production in association with three jets via vector-boson fusion: real-emission diagrams for (a) two gluons connect to the same quark line, (b) two gluons connect to different quark lines; (c) and (d) for one-loop diagrams. The black circle denotes the $HVV$ vertex.}
    \label{fig:hjjj}
\end{figure}

Figures~\ref{fig:vbf_feynman_diagrams} and \ref{fig:hjjj} summarize the perturbative contributions to vector boson fusion (VBF) Higgs production considered in this study. Figure~\ref{fig:vbf_feynman_diagrams} illustrates the Born-level process together with the representative virtual and real-emission corrections contributing to Higgs production in association with two jets ($H+2\mathrm{j}$), while Fig.~\ref{fig:hjjj} shows representative real-emission and one-loop diagrams for Higgs production in association with three jets ($H+3\mathrm{j}$). Together, these diagrams illustrate the perturbative ingredients entering the NLO calculations and the higher-multiplicity matrix elements required for multi-jet merging. They also highlight the scope of the extensions made to the \texttt{VBFNLO} framework to support anomalous $HVV$ couplings across multiple jet multiplicities. While the matrix elements for Higgs production with two jets already possessed the capability to model anomalous couplings~\cite{Figy:2003nv,Hankele:2006ma}, the $H+3\mathrm{j}$ module required significant modifications to consistently incorporate these effects at NLO accuracy, as detailed in ~\cite{Chen:2023vbf}.

VBF Higgs production provides a direct probe to both the strength and the tensor structure of the 
interactions between the Higgs boson and the electroweak gauge bosons. Taking the momenta 
$q_1$ and $q_2$ to be incoming to the $HVV$ vertex, the relevant tensor structure 
implemented in \vbfnlo can be written as~\cite{Figy:2004pt,Hankele:2006ma}:
\begin{eqnarray}
T^{\mu \nu}(q_{1},q_{2}) &=& a_1(q_1,q_2) \, g^{\mu \nu} + 
a_2(q_1,q_2) \left[q_{1}\cdot q_{2} \, g^{\mu \nu} - q_{2}^{\mu}q_{1}^{\nu}\right] 
\nonumber\\
&&+ a_3(q_1,q_2) \,
\varepsilon^{\mu \nu \rho \sigma} q_{1 \rho} q_{2 \sigma},
\label{eq:Tmunu}
\end{eqnarray}
where $\varepsilon^{\mu \nu \rho \sigma}$ is the totally antisymmetric Levi-Civita tensor 
with convention $\varepsilon_{0123}=1$. The coefficient $a_1=\frac{2m_V^2}{v}$ represents 
the SM $HVV$ couplings with $a_2=a_3=0$, while $a_2$ and $a_3$ parameterize possible 
new physics contributions. These form factors are derived from the dimension-5 effective 
Lagrangian~\cite{Figy:2004pt}:
\begin{eqnarray}
\mathcal{L}_{5} &=& 
\frac{g^{HWW}_{5e}}{\Lambda_{5e}}  HW^{+}_{\mu \nu}W^{-\mu \nu} + 
\frac{g^{HWW}_{5o}}{\Lambda_{5o}}  H \tilde{W}^{+}_{\mu \nu}W^{-\mu \nu} + 
\nonumber \\  &&
\frac{g^{HZZ}_{5e}}{2\Lambda_{5e}} HZ_{\mu \nu}Z^{\mu \nu} + 
\frac{g^{HZZ}_{5o}}{2\Lambda_{5o}} H \tilde{Z}_{\mu \nu}Z^{\mu \nu}\;,
\label{eq:Leff}
\end{eqnarray}
where the subscripts $e$ and $o$ denote CP-even and CP-odd operators, respectively, and 
the dual field-strength tensor is defined by 
$\tilde{V}_{\mu\nu}=\frac{1}{2}\epsilon_{\mu \nu \rho \sigma}V^{\rho\sigma}$ ($V=W,Z$). 
The mapping between the Lagrangian coefficients and the form factors in Eq.~\ref{eq:Tmunu} 
is given by:
\begin{equation}
a_2(q_1,q_2) = -\frac{2}{\Lambda_{5e}} g^{HWW}_{5e}\;, \qquad
a_3(q_1,q_2) = \frac{2}{\Lambda_{5o}} g^{HWW}_{5o}
\label{eq:hww}
\end{equation}
for the $HWW$ vertex, and
\begin{equation}
a_2(q_1,q_2) = -\frac{2}{\Lambda_{5e}} g^{HZZ}_{5e}\;, \qquad
a_3(q_1,q_2) = \frac{2}{\Lambda_{5o}} g^{HZZ}_{5o}
\label{eq:hzz}
\end{equation}
for the $HZZ$ vertex. We neglect possible contributions from $H\gamma\gamma$ and 
$H\gamma Z$ couplings in this study, which can appear in $SU(2)\times U(1)$ invariant 
formulations~\cite{Buchmuller:1985jz,Hagiwara:1993ck} and have only focused on the 
$HZZ$ and $HWW$ contributions.

Although we employ the dimension-5 operator formalism described above, our implementation 
is sufficiently general to accommodate Standard Model Effective Field Theory (SMEFT) 
extensions. In the SMEFT framework, dimension-6 operators generate identical tensor 
structures in the $HVV$ vertex, with the form factors $a_2$ and $a_3$ mapping linearly 
to the corresponding Wilson coefficients~\cite{Grzadkowski:2010es,Brivio:2020onw}. 
Consequently, the results presented here can be directly interpreted in terms of SMEFT 
constraints without modification of the underlying amplitude structure.

Specifically, Fig.~\ref{fig:vbf_feynman_diagrams}(a) depicts the leading-order topology 
for the $H+2\text{j}$ process. At this level, the amplitude factorizes into two quark 
currents connected by t-channel weak boson propagators and the $HVV$ tensor defined above:
\begin{equation}
\mathcal{M}_{\text{Born}} = J_\mu^{(1)}(q_1) \, D^{\mu\alpha}(q_1) \, T_{\alpha\beta}(q_1,q_2) \, 
D^{\beta\nu}(q_2) \, J_\nu^{(2)}(q_2),
\label{eq:born_factorization}
\end{equation}
where $J_\mu^{(i)}$ denotes the weak current for each quark line and $D^{\mu\nu}$ represents 
the weak boson propagator. The anomalous coupling contributions enter exclusively through 
$T_{\alpha\beta}$, allowing for a modular implementation where SM and beyond-SM 
contributions are computed separately and combined at the amplitude level.

At NLO, virtual corrections arise from gluon exchange along the quark lines 
(Fig.~\ref{fig:vbf_feynman_diagrams}(b)). Due to the color-singlet nature of the t-channel 
weak boson exchange in VBF, these corrections factorize from the electroweak structure:
\begin{equation}
\mathcal{M}_{\text{virt}} = \mathcal{M}_{\text{Born}} \times \left[ 1 + \frac{\alpha_s}{4\pi} \, 
C_F \, V(q_1,q_2) \right] + \mathcal{O}(\alpha_s^2),
\label{eq:virtual_factorization}
\end{equation}
where $V(q_1,q_2)$ encodes the loop integral structure and $C_F = 4/3$. This factorization 
ensures that the anomalous coupling dependence factors out identically in both Born and 
virtual contributions. 

Crucially, in the Deep Inelastic Scattering (DIS) approximation, this factorization 
structure extends to higher jet multiplicities. As shown in Ref.~\cite{Figy:2007kv}, 
the DIS approximation treats each quark line as an independent deep inelastic scattering 
process, with the t-channel weak bosons probing the partonic structure of the incoming 
quarks. This allows the same factorized amplitude structure to be applied to $H+3\text{j}$ 
one-loop amplitudes and $H+4\text{j}$ tree-level amplitudes, where additional jets arise 
from QCD radiation off the quark lines without disrupting the electroweak tensor structure. 
The anomalous couplings remain confined to the $HVV$ vertex regardless of jet multiplicity, 
enabling a consistent treatment across all samples used in our merging procedure.

Real-emission corrections produce an additional hard parton, contributing to the 
$H+3\text{j}$ final state (Fig.~\ref{fig:vbf_feynman_diagrams}(c)). The implementation 
of anomalous couplings in these real-emission configurations is essential for our merging 
procedure, which combines NLO-accurate matrix elements for both $H+2\text{j}$ and 
$H+3\text{j}$ multiplicities with LO-accurate $H+4\text{j}$ samples. This ensures a 
consistent description of QCD radiation effects on azimuthal observables in the presence 
of new physics.


\newcommand{\hjjcell}[1]{%
    \parbox[b][0.22\textwidth][b]{0.29\textwidth}{%
        \centering
        \includegraphics[width=0.27\textwidth]{#1}%
    }%
}
\begin{figure*}[t]
    \centering
    \setlength{\tabcolsep}{3pt}

    \begin{tabular}{@{}ccc@{}}
        \hjjcell{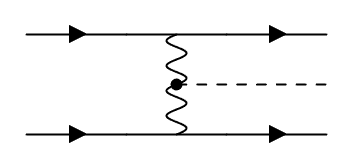}
        &
        \hjjcell{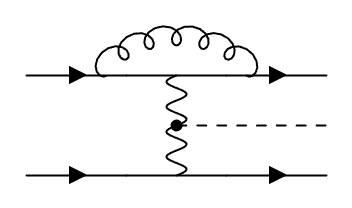}
        &
        \hjjcell{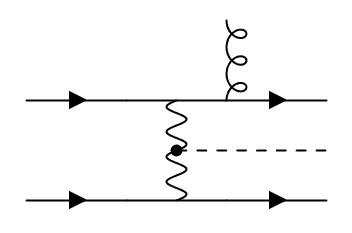}
        \\[-0.5em]
        (a) & (b) & (c)
    \end{tabular}

    \caption{Feynman diagrams for Higgs boson production through VBF. (a) the Born-level contribution, (b) virtual QCD correction involving gluon exchange along the upper quark line, and (c) real-emission contribution leading to an additional final-state jet. The enlarged filled circle denotes the $HVV$ interaction vertex.}
    \label{fig:vbf_feynman_diagrams}
\end{figure*}

\section{\label{sec:setup} Setup for Calculation}
In this section, we begin by summarizing the relevant input parameters and describing the event selection criteria employed in the analysis. For this study, simulations are performed using \texttt{Herwig 7.2.3}~\cite{Bahr:2008pv,Bellm:2015jjp,Bellm:2025pcw} as the Monte Carlo event generator, with NLO QCD matrix elements provided by \texttt{VBFNLO 3.0}~\cite{baglio2011vbfnlo,baglio2014release,Arnold:2008rz,Figy:2007kv,Baglio:2024gyp,Chen:2023vbf}. Hadronization and multiple parton interactions (MPI) are not included in the present study. All simulated events were analyzed using {\tt Rivet 2.7.2}~\cite{buckley2013rivet} toolkit.

\subsection{General Monte Carlo Input Parameters}  
To investigate the effects of anomalous Higgs couplings across different scenarios, the input parameters for \herwig closely follow the configuration detailed in Ref.~\cite{Chen:2021phj}. The masses and widths of the $Z^{0}$ and $W^{\pm}$ Gauge bosons are fixed to:

\begin{align}
    m_Z&=91.1876~\GeV, \quad \Gamma_Z=2.4952~\GeV, \\
    m_W&=80.385~\GeV, \quad \Gamma_W=2.085~\GeV.
\end{align}
The electroweak parameters are computed in the $G_\mu$ scheme with the Fermi constant set to $G_F=1.16637\times10^{-5}~\GeV^{-2}$. The Higgs boson mass is set to $m_H=125.7~\GeV$ and it is treated as a stable particle throughout the simulation. For renormalization and factorization scales, the following scale is used:
\begin{equation}
	\mu_0=\frac{1}{2} H_{T,{\rm jets}}= \frac{1}{2} \sum_{i \in \text{jets}} p_{T,i},
\end{equation}
where $p_{T,i}$ is the transverse momentum of the $i$-th jet. To ensure the infrared safety of the scale choice, jets are clustered using the anti-$k_{T}$ jet algorithm~\cite{Cacciari:2005hq, Cacciari:2008gp} with $R=0.4$ in the inclusive mode. The recombination method follows {\tt E-scheme} and each jet is required to have $p_{T,i}>5$ GeV.  All simulations are performed at a collider energy of $\sqrt{s}=13~\TeV$. The parton distribution function used is {\tt PDF4LHC15\_nnlo\_100\_pdfas}~\cite{Butterworth_2016} from {\tt LHAPDF6}~\cite{Buckley:2014ana} within all simulations. To be consistent with {\tt PDF4LHC15\_nnlo\_100\_pdfas} settings, the strong coupling constant was chosen as $\as(M_Z)=0.118$ and the five-flavor scheme is used.
\subsection{Event Generation Cuts and Kinematic Selection}
Before the subsequent parton shower, we apply a set of hard process selection cuts. The anti-kt jet algorithm is employed via the {\tt fastjet} library~\cite{Cacciari:2011ma} to reconstruct jets from final-state partons in the hard-scattering process. For the anti-kt jet algorithm~\cite{Cacciari:2005hq, Cacciari:2008gp}  we use $R=0.4$ in the inclusive mode and use the $E$-scheme for the recombination mode.  The threshold for the jet transverse momentum is $p_{T,j}>10$ GeV with the jet rapadity restricted to $|y_{j}| \geq 5$.

The analysis of events after the parton shower, we developed an analysis named {\tt MC\_H2JETS}  within the {\tt Rivet 2.7.2} framework~\cite{Buckley_2013}, which incorporates two event selection criteria: inclusive cuts (INCL) and tight cuts (TIGHT). At the analysis level, valid jets are required to satisfy certain conditions regarding their transverse momentum $p_{T,j}$ and rapidity $y_{j}$. These conditions are as follows:
\begin{equation}
p_{T,j} > 25~\GeV,\quad |y_{j}| \leq 4.5. 
\label{eq:inclusivecut}
\end{equation}
The jets are ordered from largest to smallest in jet transverse momentum and labeled jets as $j_{k}$ with $k=1,2,3,...$ being an index. For INCL selection cuts, it requires at least two jets in the event. 
For the TIGHT selection cuts,  the following additional selection criterion is included,
\begin{equation}
	m_{j_1j_2}>600~\GeV,\quad \Delta y_{j_1j_2}>4.5, \quad y_{j_1}\cdot y_{j_2}<0.
\end{equation}
\subsection{Merging and Matching in \herwig}

Matching parton showers to NLO QCD calculations has been established as the standard approach for precision simulations at hadron colliders. The term ``matching'' refers to a methodology that subtracts the $\mathcal{O}(\alpha_s)$ expansion of the parton shower from the fixed-order NLO result, thereby ensuring that physical observables and distributions after showering preserve NLO accuracy to $\mathcal{O}(\alpha_s)$ without double-counting radiative corrections~\cite{Bellm:2016rhh}. Within the matching algorithms implemented in \herwig, scale setting can be performed using various options, all of which are fully compatible with the \herwig parton showers~\cite{Platzer:2011bc,Platzer:2009jq,Gieseke:2003rz}. In the present study, we adopt the default ``resummation profile''. This profile features a narrowly smeared step function near the hard scale, ensuring a smooth transition that does not introduce unphysically small scale variations~\cite{Chen:2021phj}.

The \herwig framework provides a highly automated, comprehensive platform to interface external matrix element providers, such as \vbfnlo with the \Matchbox~\cite{Platzer:2011bc} module. \Matchbox enables the systematic assembly of fully differential NLO cross sections from these external matrix elements. Through this module, \herwig can match NLO matrix elements to both the angular-ordered shower~\cite{Gieseke:2003rz} and the dipole shower~\cite{Platzer:2009jq,Platzer:2011bc} using either a subtractive (MC@NLO-type~\cite{Frixione:2002ik}) or a multiplicative (POWHEG-type~\cite{Nason:2004rx}) matching algorithm. In our work, we focus on the dipole shower matched via the subtractive matching paradigm.

Whereas matching improves the perturbative description of the calculation, multijet merging facilitate the consistent combination of several jet multiplicities with the parton shower~\cite{Chen:2021phj}. 
The full implementation of the unitary merging algorithm in \herwig does not strictly enforce the exact preservation of the inclusive cross section\cite{Lonnblad:2011xx,Lonnblad:2012ix,Lonnblad:2012ng}; instead, subtraction terms are applied to logarithmically enhanced configurations for which a valid parton shower history can be constructed. Any configuration that cannot be clustered into a valid shower history is treated as a genuinely hard jet topology from which the parton shower evolves in a vetoed manner, preventing double-counting in both the real emission and the virtual or unresolved corrections. This approach allows the merging algorithm to natively handle processes that contain hard jets at the Born level. To smooth out threshold effects across the phase-space boundary, the merging scale $\rho_s$ is dynamically smeared on an individual event-by-event~\cite{Bellm:2017ktr}. The central merging scale is set to $\rho_{C}=25~\GeV$, and the smearing parameter is set to $\delta$ = 0.1. We adopt the CMW scheme~\cite{Catani:1990rr,Bellm:2017ktr} for the merged simulations with the modified strong coupling set to $\alpha^{\prime}_{S}(q)=\alpha_{S}(k_{g}(q))$ where $k_{g}=\exp(-K_{g}/b_{0})$, $K_g=C_A\Big(\frac{67}{18}-\frac{1}{6}\pi^2\Big)-\frac{5}{9}N_F$, and  $b_0=11-2/3N_{F}$.

\section{PHENOMENOLOGICAL RESULTS}
\label{sec:results}

In this section, we present and analyze the results of the anomalous $HVV$ couplings in the Higgs boson production process through VBF at the LHC. The discussion focuses on the impact of anomalous $HVV$ couplings on the matching and merging algorithm of parton showers. We introduce here the notation used throughout this study to describe the matching and merging algorithm of parton showers. At leading-order (LO), the Born-level hard scattering process is $p \; p \rightarrow h + j \; j$, where the corresponding tree-level $2 \rightarrow h+2$ partonic matrix elements (MEs) are denoted by $h(2)$. For LO multijet merging setups, additional tree-level MEs with higher parton multiplicities are included, and the notation $h(2,3,\dots, n)$ is used to indicate merged MEs for $h+2$, $h+3$, up to $h+n$ partons, with $n\geq 2$. To denote the inclusion of NLO corrections, we append an asterisk (e.g. $n^\star$) to any multiplicity $n$, indicating that both tree-level and one-loop MEs are used for that multiplicity. For instance, \hjjmerge refers to a merged sample where one-loop corrections are included for the $h+2$ and $h+3$ parton processes, while tree-level matrix elements are used for the $h+4$ processes. A special case is \hjjmerge[3] which represents the “matching through merging” configuration employed by \herwig, achieving the same formal accuracy as standard NLO matching \cite{Chen:2021phj}. In cases where subtractive NLO matching is used to couple fixed-order NLO calculations with a parton shower (e.g., the dipole shower), the notation \hjjnlops is used for the $p \; p \rightarrow h + j \; j$ process, indicating that NLO MEs for the corresponding multiplicity are matched to a parton shower. We also denote that the \hjjlops for the LO MEs matching with the dipole shower for the Higgs boson plus two jets production process.

\subsection{CP Sensitive Observables}
Throughout this section, the anomalous couplings are set in three scenarios:
\begin{equation}
\begin{aligned}
    \text{Pure CP-even}: \quad & g_{5e}^{HWW} = g_{5e}^{HZZ} = 0.5, \\ 
    \text{Pure CP-odd}:  \quad & g_{5o}^{HWW} = g_{5o}^{HZZ} = 0.5, \\ 
    \text{CP-mixed}:     \quad & g_{5e}^{HWW} = g_{5o}^{HWW} = g_{5e}^{HZZ} = g_{5o}^{HZZ} = 0.5.
\end{aligned}
\label{eq:cp-case}
\end{equation}
In all three CP scenarios considered, the SM contribution is excluded by setting $a_1=0$ in Eq.~\ref{eq:Tmunu}, with the $\Lambda_{5e}=\Lambda_{5o}=480$\GeV. 
The fiducial cross sections for the SM, CP-even, CP-odd, and CP-mixed scenarios are summarized in Table~\ref{tab:cross_sections}. For each coupling scenario, results are presented after applying the INCL and TIGHT selection cuts using the different setup: NLO matched parton shower \hjjnlops, \hjjmerge[3], and the multijet merging \hjjmerge. The uncertainties correspond to the Monte Carlo integration uncertainties.
The CP properties of the $HVV$ vertex can be probed by examining the azimuthal angle separation between the two leading jets~\cite{Figy:2004pt,Hankele:2006ma}. In general, this angle is defined as the azimuthal difference between the two hardest jets,
\begin{equation}
    |\Delta\phi_{j_1j_2}|=|\phi_{j_1}-\phi_{j_2}|.
    \label{eq:deltaphi}
\end{equation}
\begin{table*}[t]
\centering
\small
\caption{
Fiducial cross sections in pb for the SM, CP-even, CP-odd, and CP-mixed scenarios after the INCL and TIGHT VBF selection cuts.
}
\label{tab:cross_sections}
\renewcommand{\arraystretch}{1.2}
\setlength{\tabcolsep}{2pt}
\resizebox{\textwidth}{!}{%
\begin{tabular}{llcccc}
\hline
\hline
\textbf{Selection Cuts} 
& \textbf{Scenario}
& \textbf{\hjjnlops}
& \textbf{\hjjmerge[3]}
& \textbf{\hjjmerge} \\ 
\hline

\multirow{3}{*}{INCL}
& SM
& $2.289032 \pm 5.864127 \times 10^{-4}$ 
& $2.319501\pm 1.284643 \times 10^{-3}$ 
& $2.324619 \pm 1.010710 \times 10 ^{-3}$ \\

& CP-even
& $7.504352\times 10^{-1} \pm 9.380646 \times 10^{-5}$  
& $7.604368 \times 10^{-1} \pm 2.458142 \times 10^{-4}$
&$7.621666 \times 10^{-1} \pm 1.381144\times 10^{-3}$\\

& CP-odd
& $6.110633 \times 10^{-1} \pm 9.134671\times 10^{-5}$
& $6.191370 \times 10^{-1}\pm 1.907430 \times 10^{-4}$    
& $6.203703 \times 10^{-1}\pm 1.137503\times 10^{-3}$ \\

& CP-mixed
& $1.361444 \pm 2.967522 \times 10^{-4}$ 
& $1.380040\pm 4.825101 \times 10^{-4}$
& $1.380876 \pm 2.925703\times 10^{-3}$ \\

\hline

\multirow{3}{*}{TIGHT}
& SM
& $8.127734\times 10^{-1}\pm 1.743162 \times 10^{-5}$ 
& $8.102169\times 10^{-1}\pm 2.677023 \times 10^{-4}$ 
& $8.121349\times 10^{-1}\pm 4.011864\times 10^{-4}$ \\

& CP-even
& $7.878944\times 10^{-2} \pm 2.436551\times 10^{-5}$ 
& $7.934439\times 10^{-2}\pm 4.483174\times 10^{-5}$ 
& $7.945200\times 10^{-2} \pm 1.398008 \times 10^{-4}$ \\

& CP-odd
& $7.427244 \times 10^{-2}\pm 1.743162 \times 10^{-5}$ 
& $7.472956\times 10^{-2}\pm 3.851067\times 10^{-5}$  
& $7.505319\times 10^{-2}\pm 1.729061\times 10^{-4}$ \\

& CP-mixed
& $1.530876\times 10^{-1} \pm 3.919667\times 10^{-5}$ 
& $1.541077\times 10^{-1} \pm 7.974181\times 10^{-5}$  
& $1.545555\times 10^{-1}\pm 2.942562\times 10^{-4}$ \\

\hline
\hline
\end{tabular}
}
\end{table*}

\begin{figure*}[!tp]
  \centering

  \begin{subfigure}[t]{0.48\textwidth}
    \includegraphics[width=0.9\textwidth]{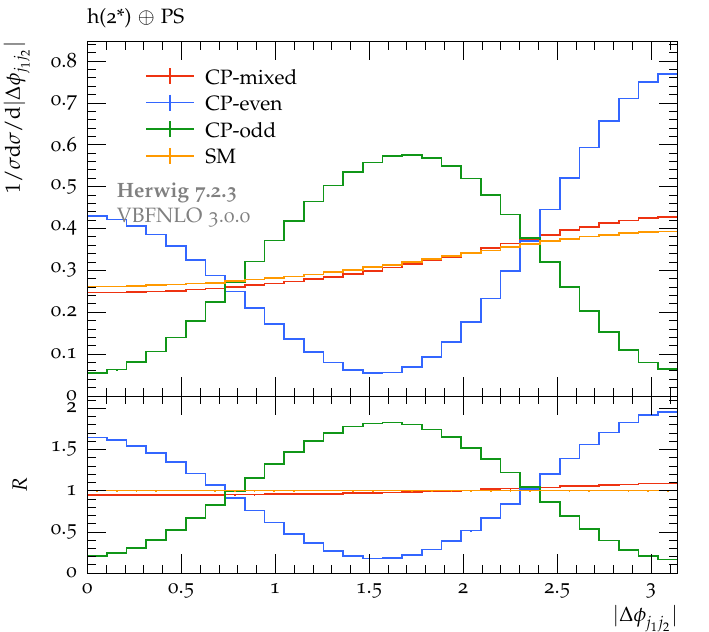}
  \end{subfigure}
  \hfill
  \begin{subfigure}[t]{0.48\textwidth}
    \includegraphics[width=0.9\textwidth]{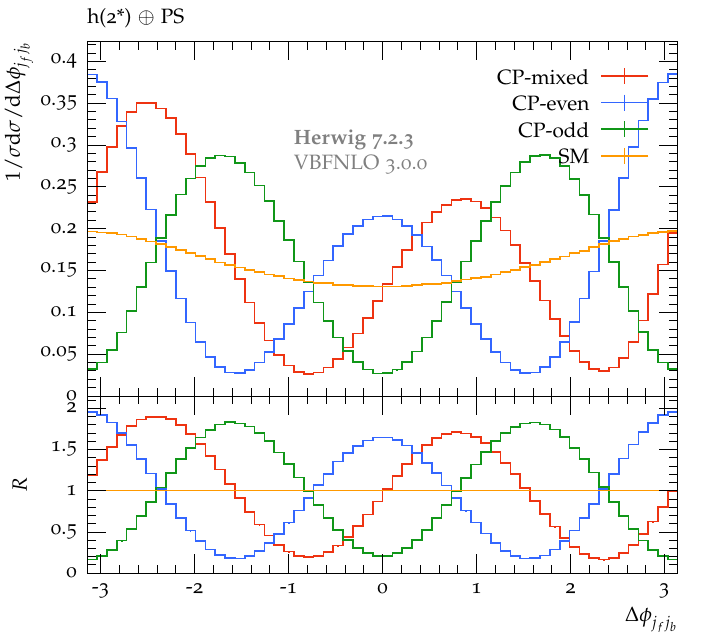}
  \end{subfigure}


  \begin{subfigure}[t]{0.48\textwidth}
    \includegraphics[width=0.9\textwidth]{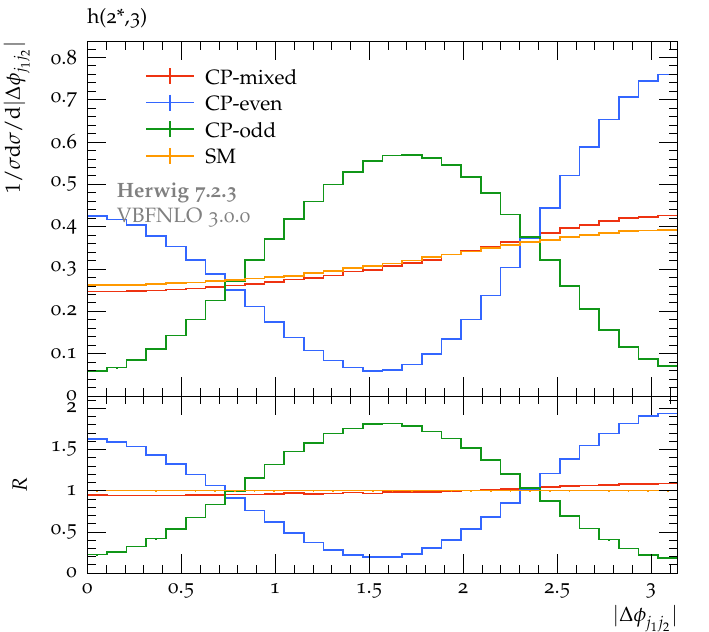}
  \end{subfigure}
  \hfill
  \begin{subfigure}[t]{0.48\textwidth}
    \includegraphics[width=0.9\textwidth]{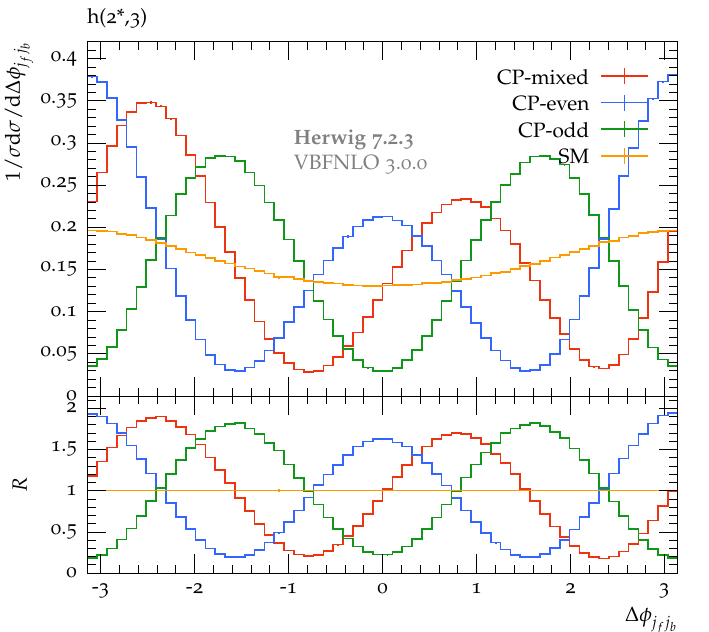}
  \end{subfigure}


  \begin{subfigure}[t]{0.48\textwidth}
    \includegraphics[width=0.9\textwidth]{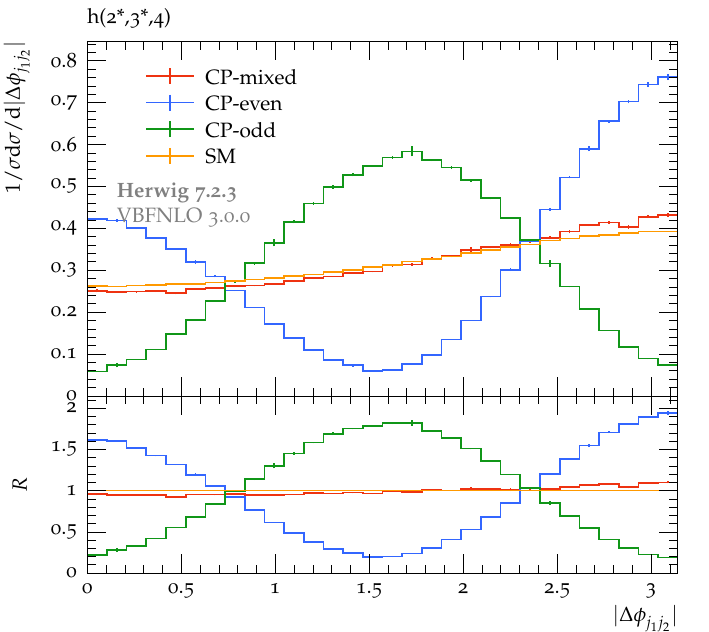}
  \end{subfigure}
  \hfill
  \begin{subfigure}[t]{0.48\textwidth}
    \includegraphics[width=0.9\textwidth]{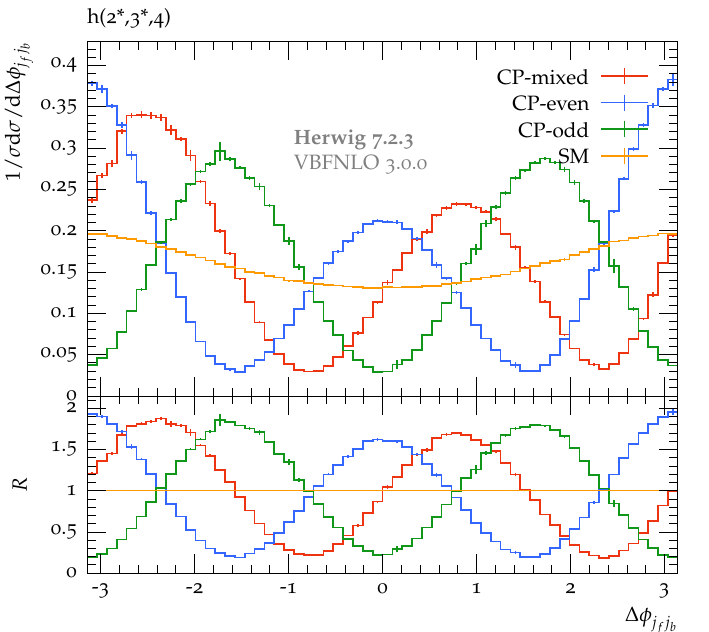}
  \end{subfigure}

  \caption{Normalized distribution of the unsigned azimuthal angle difference $|\Delta\phi_{j_1j_2}|$ (left) and signed azimuthal angle difference between the forward and backward tagging jets $\Delta\phi_{j_fj_b}$ (right) for the CP-mixed, CP-even, CP-mixed, and SM scenarios. The plots show predictions obtained with matched setup \hjjnlops(top), \hjjmerge[3](middle), and merged setup \hjjmerge(bottom) using TIGHT selection cuts. The lower panels display the ratio $R$ of the anomalous coupling predictions against the SM results.}
  \label{fig:cp_deltaphi}
\end{figure*}

A direct comparison of the \dphijj among the different anomalous coupling scenarios against the SM prediction is shown in the left column of Fig.~\ref{fig:cp_deltaphi}. The error bars in each figure represent Monte Carlo statistical uncertainties. The bottom panel of each plot shows the factor of $R$, which is the ratio of normalized distributions relative to the SM prediction, defined as
\begin{equation}
    R(X)=\frac{1/\sigma^{CP} d\sigma^{CP}/ dX}{1/\sigma^{SM}d\sigma^{SM}/dX},
    \label{eq:r}
\end{equation}
where $X$ represents the observable (i.e., \dphijj). The superscripts $SM$ and $CP$ represent the SM and three CP scenarios shown in Eq.~\ref{eq:cp-case}.
For the purely CP-even scenario, the normalized differential cross section of \dphijj exhibits a suppression around 90 degrees, whereas in the purely CP-odd case, suppression occurs near 0 and 180 degrees. This behavior is observed in both matched setups \hjjnlops and merged setups, \hjjmerge[3], and \hjjmerge, where the CP-odd case shows a peak at 90 degrees, while the CP-even case shows a pronounced dip. In the CP-mixed scenario, where CP-even and CP-odd contributions are of equal magnitude, the distinctive features of each cancel out, resulting in a relatively flat distribution and remaining a valid discriminant when comparing purely CP-even and CP-odd anomalous couplings. The normalized distribution of CP-mixed shows about $10\%$ deviation at $|\Delta\phi_{j_1j_2}|=\pi$ for all setups.
Since Eq.\ref{eq:deltaphi} involves the absolute difference between $\phi_{j_1}$ and $\phi_{j_2}$, it inherently discards the sign information of the azimuthal separation between the tagging jets. By identifying the ``forward" and ``backward" jets relative to the beam and defining \dphijfjb as the azimuthal angle of the ``away" jet minus the ``toward" jet, the sign of \dphijfjb remains invariant under beam direction exchange~\cite{Hankele:2006ma}. \dphijfjb is defined as 
\begin{align}
    \Delta\phi_{j_fj_b}=\begin{cases}
        \phi_{j_1}-\phi_{j_2},\quad y_{j_1}>y_{j_2}\\
        \phi_{j_2}-\phi_{j_1},\quad y_{j_1}<y_{j_2}.
    \end{cases}
\end{align}
In order to make $\Delta\phi_{j_fj_b}\in[-\pi,\pi]$, $\Delta\phi_{j_fj_b}$ also satisfies
\begin{align}
    \Delta\phi_{j_fj_b}=\begin{cases}
        \Delta\phi_{j_fj_b}+2\pi,\quad \Delta\phi_{j_fj_b}<-\pi\\
        \Delta\phi_{j_fj_b}-2\pi,\quad \Delta\phi_{j_fj_b}>\pi.
    \end{cases}
\end{align}
 
The normalized distributions for \dphijfjb under TIGHT selection cuts are presented in the right column of Fig.~\ref{fig:cp_deltaphi} for \hjjnlops, \hjjmerge[3], and \hjjmerge, comparing three anomalous coupling scenarios against the SM case. The CP-even and CP-odd cases exhibit qualitative behavior similar to that observed in \dphijj as $\Delta\phi_{j_fj_b}\in [0,\pi]$ in all matched and merged setups. However, for mixed CP couplings, CP-even and CP-odd contributions do not cancel completely, instead the maxima shift to $\Delta\phi_{j_fj_b} = \frac{\pi}{4}$ and $-\frac{3\pi}{4}$, while the minima occur at $\frac{3\pi}{4}$ and $-\frac{\pi}{4}$. The shift reflects the missing features in the distribution of \dphijj. The predictions of the matched and merged setup exhibit similar behavior, with the case of mixed CP showing shifted dip locations when the sign of $\Delta\phi_{j_fj_b}$ is retained.

\begin{figure*}[!tp]
  \centering
  \includegraphics[width=0.432\textwidth]{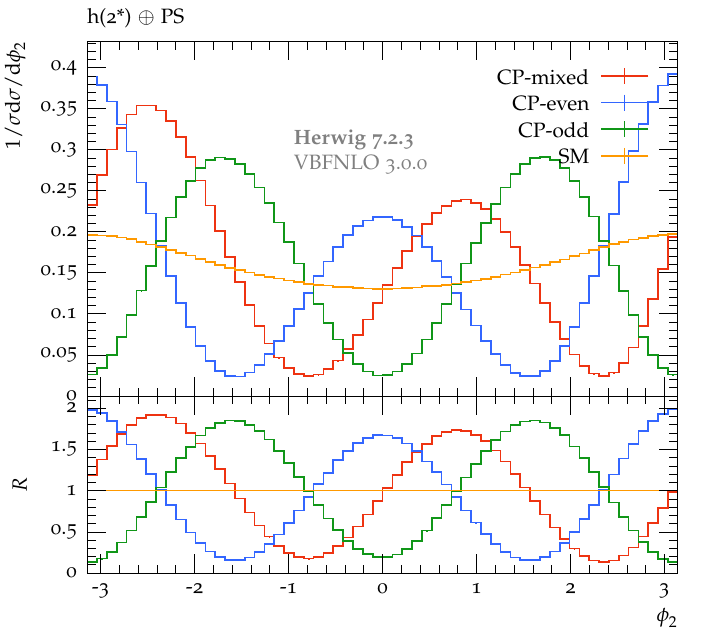}
  \includegraphics[width=0.432\textwidth]{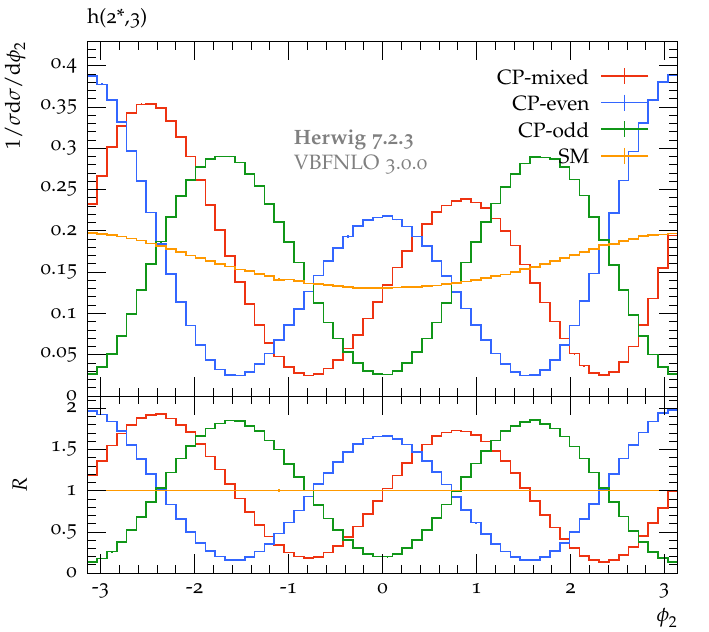}
  \includegraphics[width=0.432\textwidth]{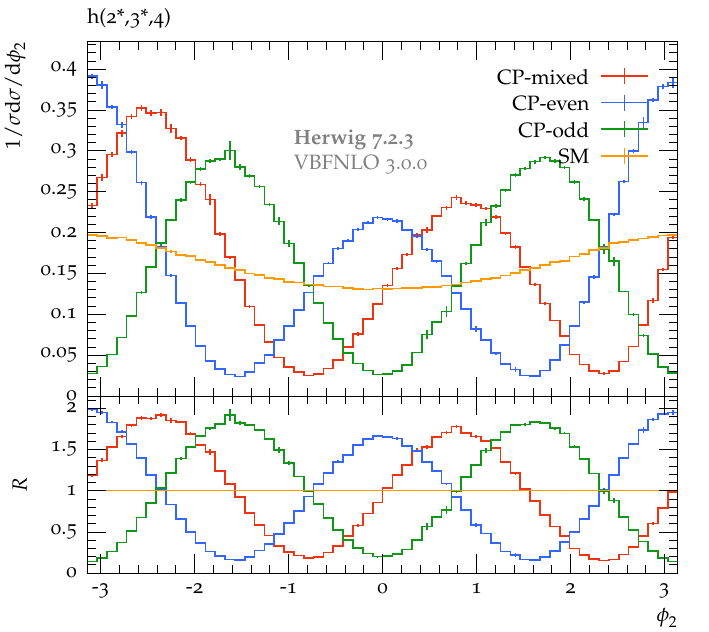}
  \caption{Normalized distributions of $\phi_2$ for CP-even, CP-odd, CP-mixed and SM scenarios, using matched setup \hjjnlops(top left), \hjjmerge[3](top right), and merged setup \hjjmerge(bottom) under TIGHT selection cuts. The lower panels show the ratio $R$ of the anomalous coupling predictions compared to the corresponding SM prediction.}
  \label{fig:cp_phi2}
\end{figure*}

While Fig.~\ref{fig:cp_deltaphi} demonstrates that the CP structure of the anomalous $HVV$ couplings can be probed using the modified azimuthal angle $\Delta\phi_{j_fj_b}$ between the two jets, an alternative observable is the $\phi_2$~\cite{Andersen:2010zx}, defined between the incoming quark momenta $\mathbf{q}_a$ and $\mathbf{q}_b$ as,
\begin{equation}
    \phi_2=\angle(\mathbf{q}_{a\perp},\mathbf{q}_{b\perp}),
\end{equation}
where
\begin{align}
    q_a=\sum_{j\in\{\text{jets}:y_j<y_h\}} p_j, \quad q_b=\sum_{j\in\{\text{jets}:y_j>y_h\}} p_j, 
\end{align}
We categorize the jets into two groups based on their rapidity relative to the Higgs boson. $\mathbf{q}_{a\perp}$ and $\mathbf{q}_{b\perp}$ are transverse components of $\mathbf{q}_{a}$ and $\mathbf{q}_{b}$. Due to the rapidity and invariant mass requirements imposed by the TIGHT selection cuts, the $\phi_2$ distribution is expected to show a similar pattern of $\Delta\phi_{j_fj_b}$. The $\phi_2$ distributions under TIGHT selection cuts are presented in Fig.~\ref{fig:cp_phi2} for \hjjnlops, \hjjmerge[3], and \hjjmerge. These results show a sinusoidal distribution similar to \dphijfjb. The CP-mixed case produces a characteristic curve shifted by $\frac{\pi}{4}$ relative to either the pure CP-even or CP-odd cases, and provides clear discrimination between the different CP structures.

\subsection{Comparison Between Matching and Merging Results}

The differences between the merged and matched setups are presented in this section. We have compared the merged configurations \hjjmerge, the matched setup \hjjlops, \hjjnlops, and the fixed-order NLO calculation $h(2^\star)$. Here we define the renormalization scale to be $\mu_R = \xi_R\mu_0$ and the factorization scale to be $\mu_F = \xi_F \mu_0$ with $\xi_F$ and $\xi_R$ denoting the scale factors. The following figures in this section show that the error bands are due to the variation of the renormalization and factorization scale factors $\xi_F$ and $\xi_R$ with $\xi_F = \xi_R = \frac{1}{2}, 1$, and $2$. To assess the impact of the NLO merging results, we compare LO and NLO results by plotting the dynamical $K$ factor~\cite{Figy:2008zd},
\begin{equation}
    K(X)=\frac{ d\sigma^{NLO_i}/ dX}{d\sigma^{LO\oplus PS}/dX},
    \label{eq:k}
\end{equation}
where $X$ is observable, $i$ represents the fixed-order NLO calculation $h(2^\star)$, matched setup $h(2^\star)\oplus\text{PS}$, and merged setup $h(2^\star,3^\star,4)$. 

\begin{figure*}[!tp]
  \centering

  \begin{subfigure}[t]{0.48\textwidth}
    \includegraphics[width=0.9\textwidth]{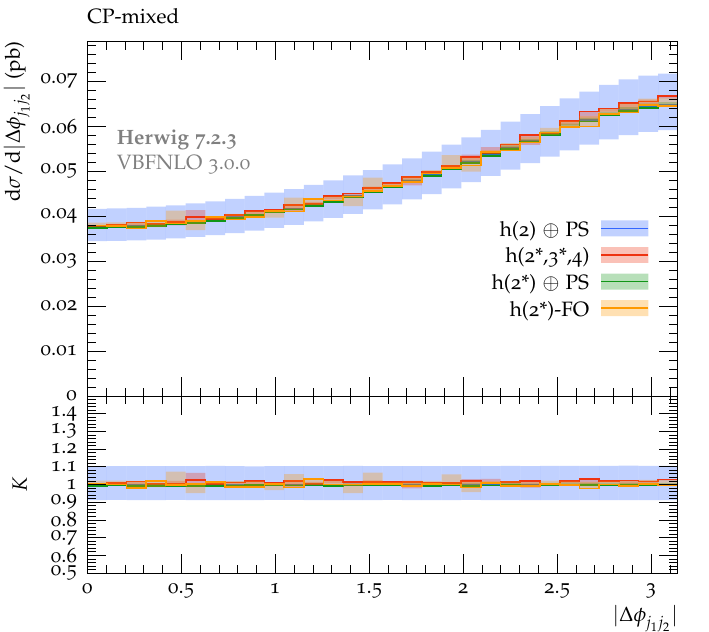}
  \end{subfigure}
  \hfill
  \begin{subfigure}[t]{0.48\textwidth}
    \includegraphics[width=0.9\textwidth]{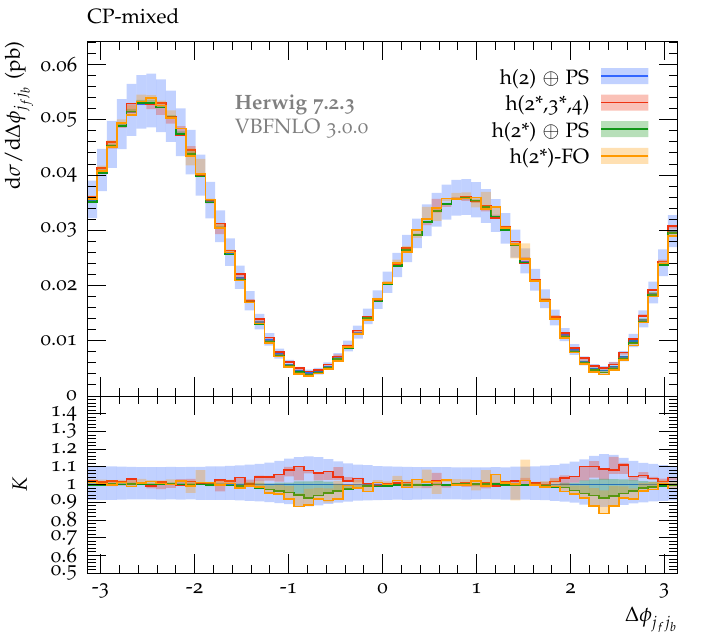}
  \end{subfigure}


  \begin{subfigure}[t]{0.48\textwidth}
    \includegraphics[width=0.9\textwidth]{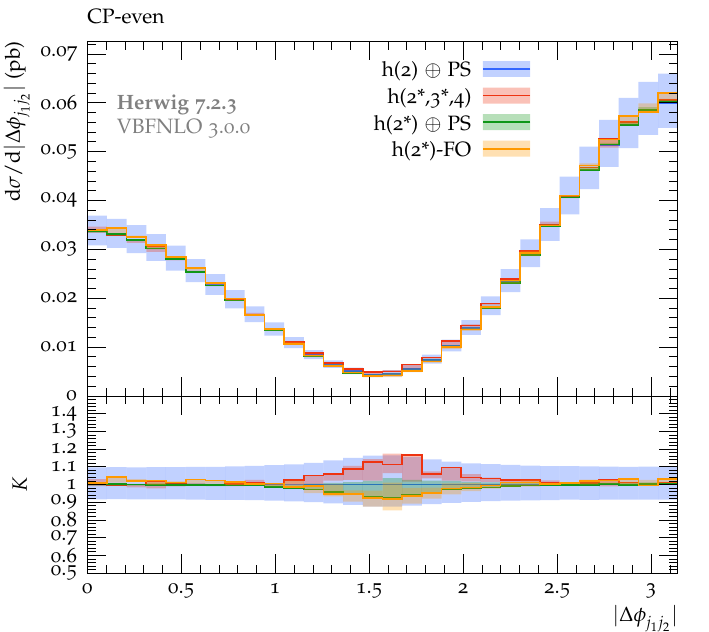}
  \end{subfigure}
  \hfill
  \begin{subfigure}[t]{0.48\textwidth}
    \includegraphics[width=0.9\textwidth]{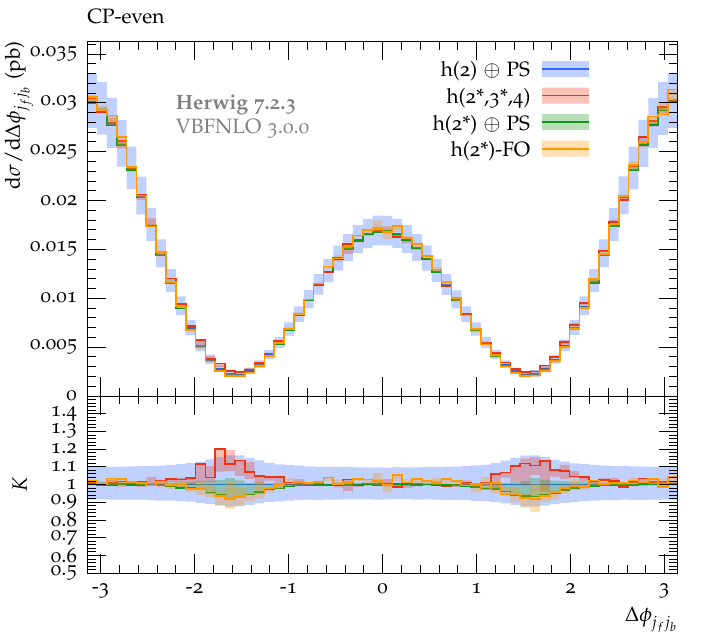}
  \end{subfigure}


  \begin{subfigure}[t]{0.48\textwidth}
    \includegraphics[width=0.9\textwidth]{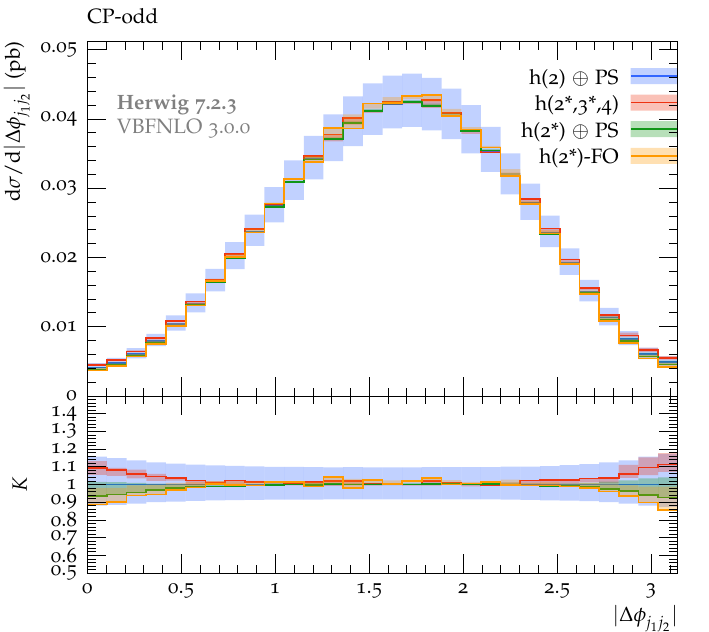}
  \end{subfigure}
  \hfill
  \begin{subfigure}[t]{0.48\textwidth}
    \includegraphics[width=0.9\textwidth]{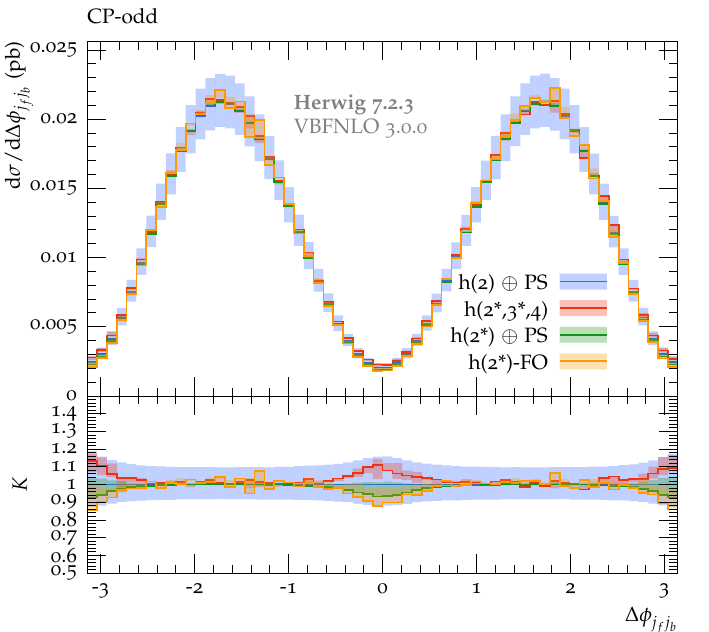}
  \end{subfigure}

  \caption{Distribution of the unsigned azimuthal angle difference $|\Delta\phi_{j_1j_2}|$ (left) and the signed azimuthal angle difference $\Delta\phi_{j_fj_b}$ (right) between the forward and backward tagging jets for the different anomalous coupling scenarios: CP-mixed (top), CP-even (middle), and CP-odd (bottom) using TIGHT selection cuts. The predictions obtained with the matched \hjjnlops and \hjjlops calculations are compared with the merged \hjjmerge prediction and the fixed-order \hjjnlo result. The lower panels show the ratios, $K$, with respect to the \hjjlops prediction. The shaded bands represent the corresponding theoretical uncertainties.}
  \label{fig:B_deltaphi_error}
\end{figure*}

\begin{figure*}[!tp]
  \centering
  \includegraphics[width=0.432\textwidth]{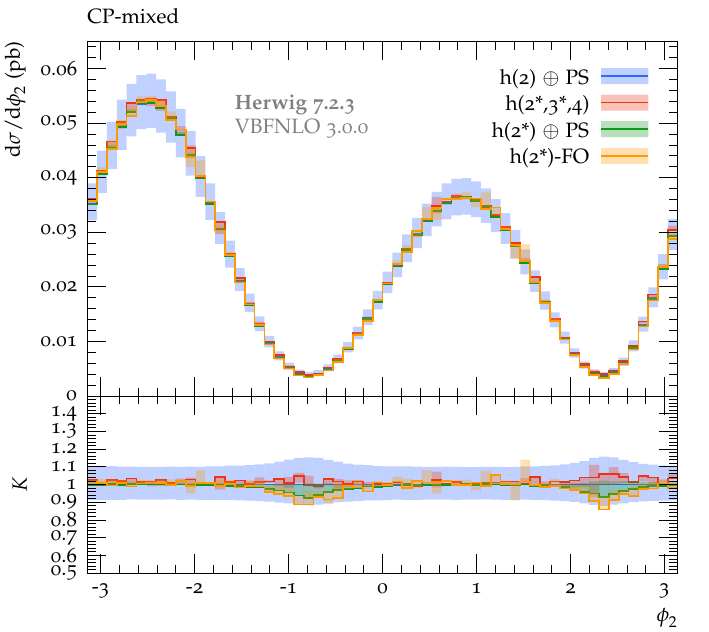}
  
  \includegraphics[width=0.432\textwidth]{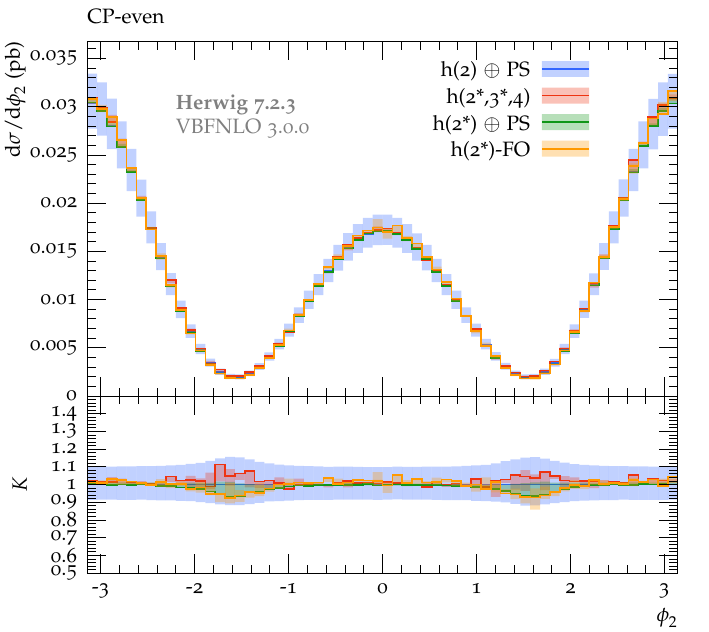}
  \includegraphics[width=0.432\textwidth]{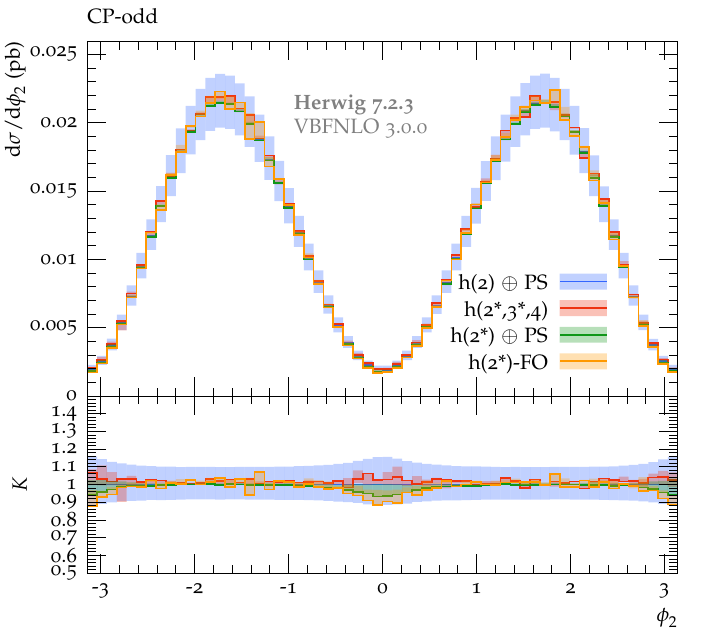}
  
  \caption{Distribution of $\phi_2$ for CP-mixed (top), CP-even (left), and CP-odd (right) cases, obtained
with the matched \hjjlops and \hjjnlops setup, the \hjjmerge merging setup, and the fixed order \hjjnlo calculation with TIGHT cuts. The lower panels display the ratios to the LO matched prediction \hjjlops, with the error bands indicating theoretical uncertainties.}
  \label{fig:B_phi2_error}
\end{figure*}

Distributions of the unsigned azimuthal angle difference \dphijj and the signed azimuthal angle difference between tagging jets, \dphijfjb, are presented in Fig.~\ref{fig:B_deltaphi_error} for the CP-mixed, CP-even, and CP-odd scenarios under TIGHT selection cuts. In the CP-mixed scenario (top row in Fig.~\ref{fig:B_deltaphi_error}), the fixed-order calculation $h(2^\star)$ and the matched setup \hjjnlops generally agree with the merged setup \hjjmerge in the \dphijj distribution, exhibiting fluctuations in regions of lower statistics. Deviations around $20\%$ between \hjjmerge and \hjjnlo are noticeable near $\Delta\phi_{j_fj_b}=-\frac{\pi}{4}$ and $\frac{3\pi}{4}$. All alternative setups remain within the error band of the LO matched prediction \hjjlops throughout the full kinematic range.
The middle row of Fig.~\ref{fig:B_deltaphi_error} presents the distributions of \dphijj and \dphijfjb for the CP-even scenario including renormalization and factorization uncertainties. For the $|\Delta\phi_{j_1j_2}|$ distribution shown in the top panel, all setups exhibit the characteristic CP-even behavior with a pronounced dip around $|\Delta\phi_{j_1j_2}|= \frac{\pi}{2}$. The fixed-order prediction \hjjnlo falls below the merged result by approximately $20\%$ around $\frac{\pi}{2}$, while the LO matched setup \hjjlops shows larger deviations through the ranges. Away from the dip region, all other predictions remain compatible within the uncertainties. The CP-even structure produces symmetric minima around $\Delta\phi_{j_fj_b}=\pm\frac{\pi}{2}$ and maxima near $\Delta\phi_{j_fj_b}=\pm\pi$. Similar to the unsigned distribution, the fixed-order predictions differ from the merged setup by up to approximately $20\%$ around $\Delta\phi_{j_fj_b}=\pm \frac{\pi}{2}$. 
The CP-odd scenario exhibits a complementary behavior compared to the CP-even distributions, with the \dphijj distribution peaking near $\frac{\pi}{2}$ and the signed distribution \dphijfjb developing pronounced maxima around $\pm\frac{\pi}{2}$. All simulation setups reproduce these characteristic CP-odd features consistently. The differences between the merged \hjjmerge, matched \hjjnlops, and fixed-order calculations remain below approximately $20\%$ and are fully covered by the uncertainty band of the LO matched setup \hjjlops. Overall, the azimuthal angle observables remain stable against higher-order corrections and parton-shower effects, confirming their robustness as probes of the CP structure of the $HVV$ interaction. The results demonstrate that the merged setup provides stable predictions for CP-sensitive azimuthal observables while maintaining consistency with the matched and fixed-order calculations. 
Fig.~\ref{fig:B_phi2_error} shows the distributions of the observable $\phi_2$ for the CP-mixed, CP-even, and CP-odd scenarios. The CP-mixed scenario produces an intermediate oscillatory structure resulting from the interference between CP-even and CP-odd contributions. The CP-even distribution exhibits minima around $\phi_2= \pm\frac{\pi}{2}$, while the CP-odd case shows the opposite behavior, with pronounced peaks in the same regions. Across all three scenarios, the merged setup \hjjmerge, the NLO matched setup \hjjnlops, and the fixed-order calculation \hjjnlo remain within the error band of the LO matched setup \hjjlops. $20\%$ deviations are visible near the minima of the distributions between \hjjmerge and \hjjnlo. These results indicate that the $\phi_2$ observable preserves its sensitivity to the CP structure of the $HVV$ interaction.

\begin{figure*}[!tp]

  \centering

  \begin{subfigure}[t]{0.48\textwidth}
    \centering
    \includegraphics[width=0.9\textwidth]{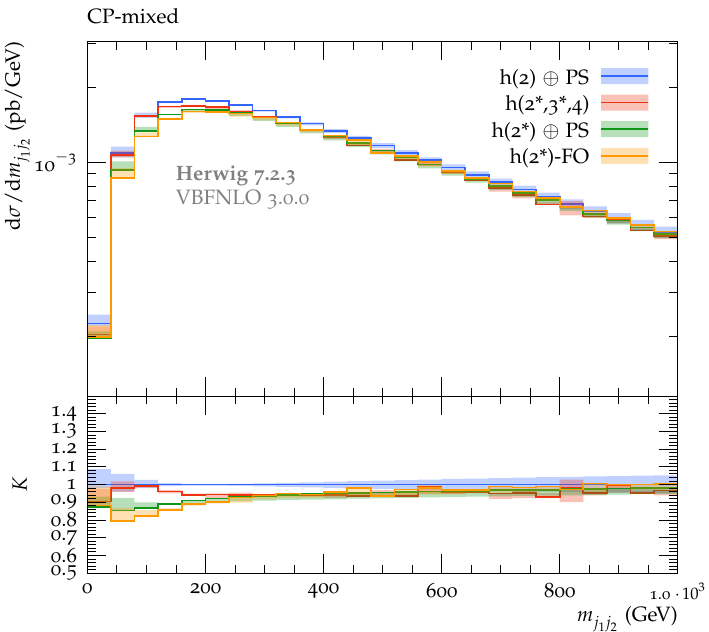}
  \end{subfigure}
  \hfill
  \begin{subfigure}[t]{0.48\textwidth}
    \centering
    \includegraphics[width=0.9\textwidth]{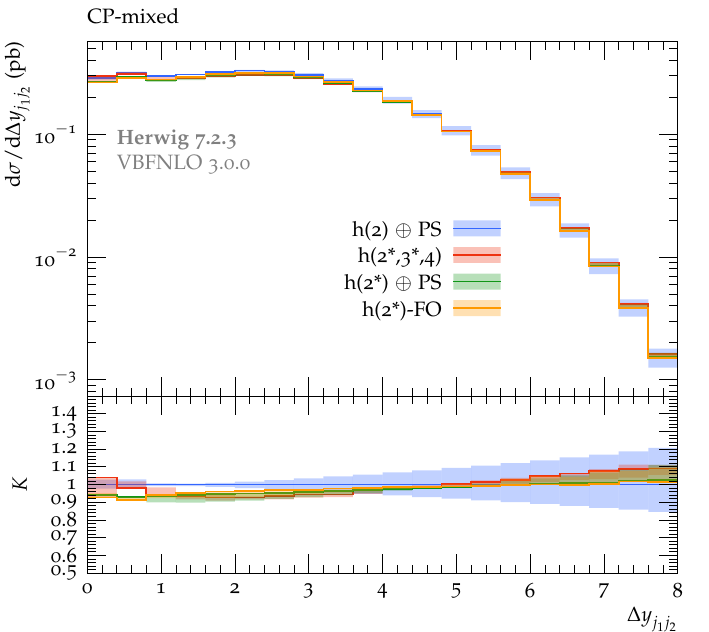}

  \end{subfigure}


  \begin{subfigure}[t]{0.48\textwidth}
    \centering
    \includegraphics[width=0.9\textwidth]{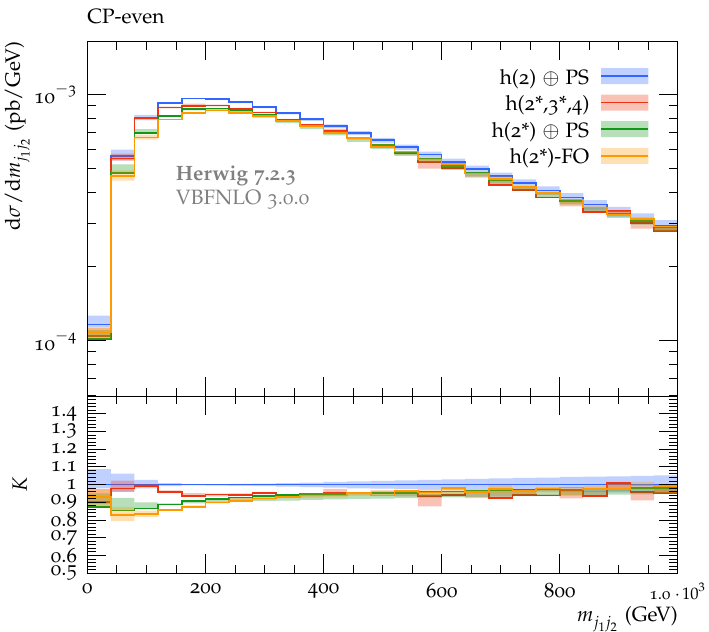}

  \end{subfigure}
  \hfill
  \begin{subfigure}[t]{0.48\textwidth}
    \centering
    \includegraphics[width=0.9\textwidth]{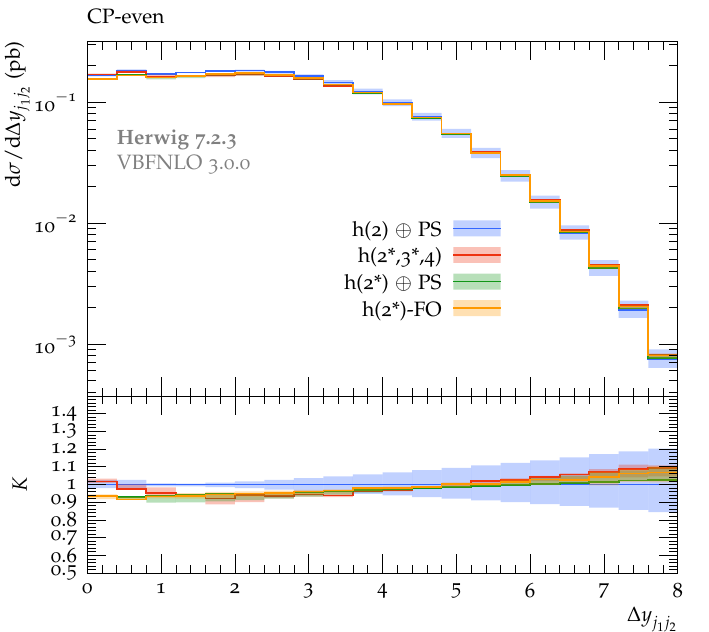}
  \end{subfigure}


  \begin{subfigure}[t]{0.48\textwidth}
    \centering
    \includegraphics[width=0.9\textwidth]{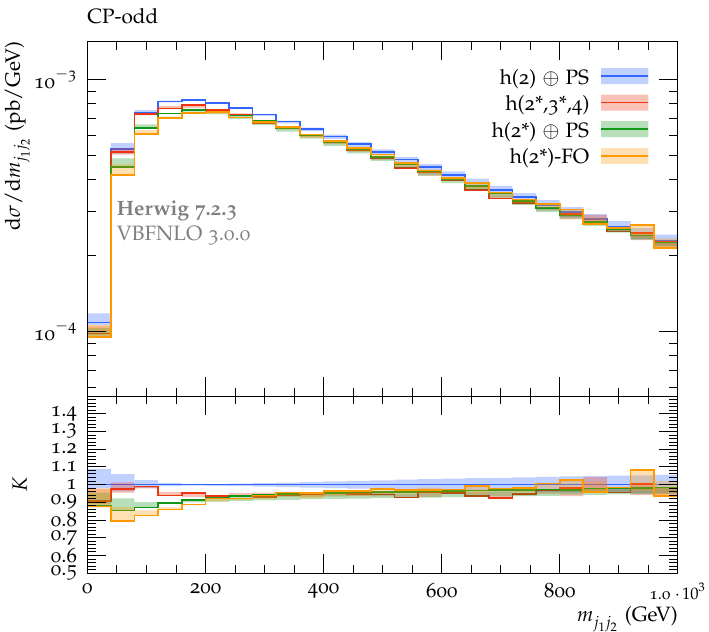}

  \end{subfigure}
  \hfill
  \begin{subfigure}[t]{0.48\textwidth}
    \centering
    \includegraphics[width=0.9\textwidth]{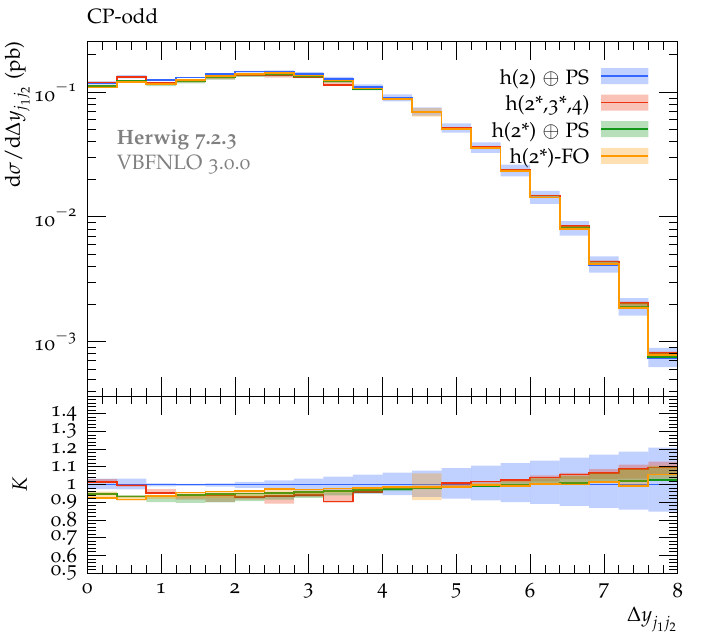}
  \end{subfigure}

  \caption{Distributions of the invariant mass of the two leading jets, $m_{j_1j_2}$ (left), and the rapidity separation between the two leading jets, $\Delta y_{j_1j_2}$ (right), for the CP-mixed (top), CP-even (middle), and CP-odd (bottom) scenarios under INCL selection cuts. Predictions obtained from the LO matched setup \hjjlops, the merged setup \hjjmerge, the NLO matched setup \hjjnlops, and the fixed-order calculation \hjjnlo are compared. The lower panels show the ratio $K$ with respect to the LO matched prediction, with the error bands indicating the corresponding theoretical uncertainties.}
  \label{fig:B_dijet_error}
\end{figure*}

Fig.~\ref{fig:B_dijet_error} presents the invariant mass distribution of the two leading jets, \mjj, and the rapidity separation distribution, \dyjj for the CP-even, CP-mixed, and CP-odd scenarios, including the renormalization and factorization uncertainties. The overall shapes of the distributions are similar across the three CP configurations under INCL selection cuts, with moderate differences. For the invariant mass distribution, three CP scenarios show a similar trend: at the lower invariant masses $m_{j_1j_2}<200~\text{GeV}$ region, the \hjjnlops and fixed-order \hjjnlo distributions deviate from the \hjjmerge prediction up to $12\%$. Beyond the region where $m_{j_1j_2}>200~\text{GeV}$, three setups exhibit agreement within statistical uncertainties. 
A similar behavior is observed in the rapidity separation distributions. The merged setup \hjjmerge, the NLO matched setup \hjjnlops, and the fixed-order calculation \hjjnlo provide consistent predictions within the theoretical uncertainty across most of the range except at low rapidity separations $\Delta y_{j_1 j_2}<2$. The fluctuations occur at $\Delta y_{j_1j_2}<2$, the \hjjnlops and fixed-order \hjjnlo results differ from the \hjjmerge prediction by roughly $10\%$, where the statistical uncertainties become sizable due to limited event statistics. In three anomalous coupling cases,  all predictions remain compatible within the uncertainty envelope of the LO matched setup in the high rapidity separation region. 

\begin{figure*}[!tp]

  \centering

  \begin{subfigure}[t]{0.48\textwidth}
    \centering
    \includegraphics[width=0.9\textwidth]{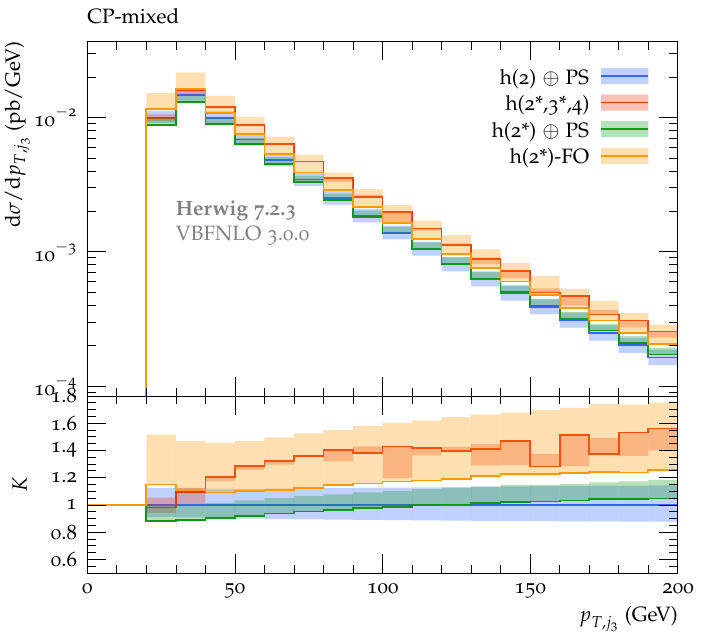}
  \end{subfigure}
  \hfill
  \begin{subfigure}[t]{0.48\textwidth}
    \centering
    \includegraphics[width=0.9\textwidth]{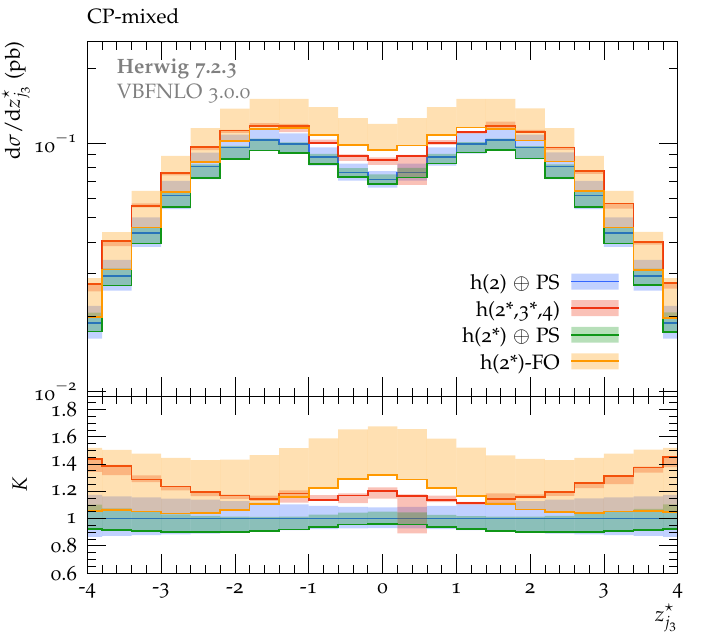}

  \end{subfigure}


  \begin{subfigure}[t]{0.48\textwidth}
    \centering
    \includegraphics[width=0.9\textwidth]{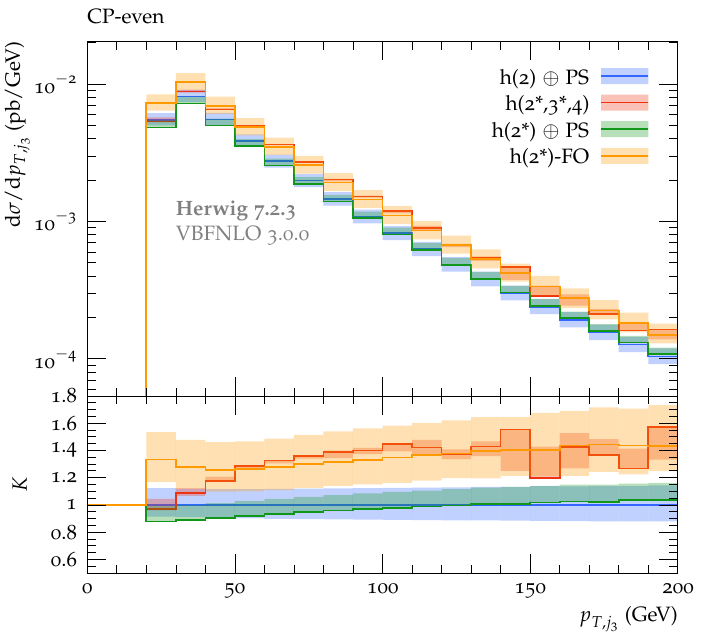}

  \end{subfigure}
  \hfill
  \begin{subfigure}[t]{0.48\textwidth}
    \centering
    \includegraphics[width=0.9\textwidth]{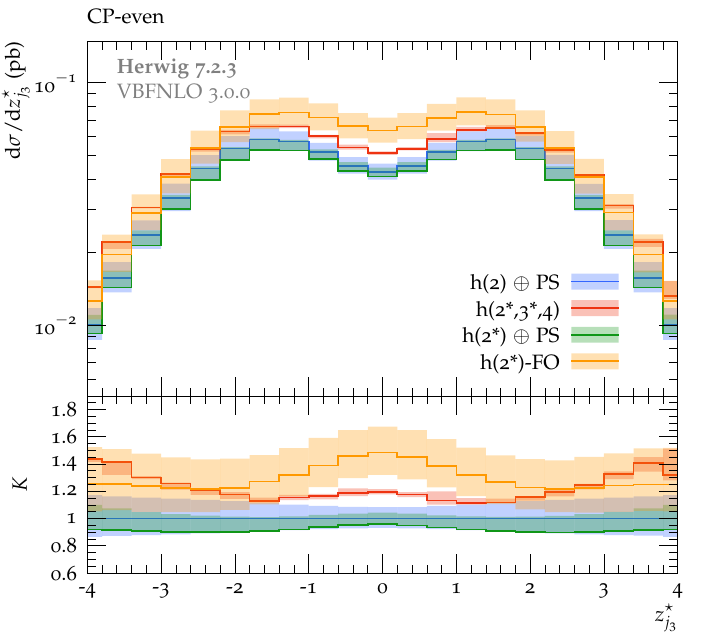}
  \end{subfigure}


  \begin{subfigure}[t]{0.48\textwidth}
    \centering
    \includegraphics[width=0.9\textwidth]{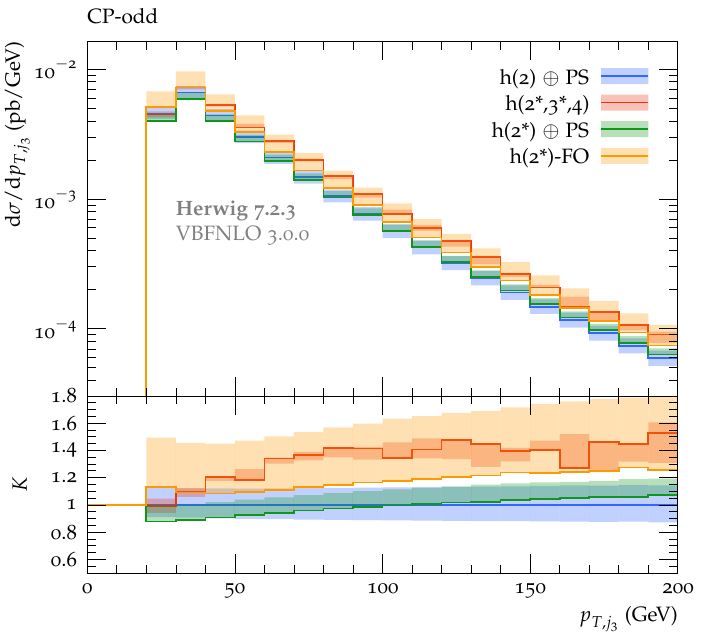}

  \end{subfigure}
  \hfill
  \begin{subfigure}[t]{0.48\textwidth}
    \centering
    \includegraphics[width=0.9\textwidth]{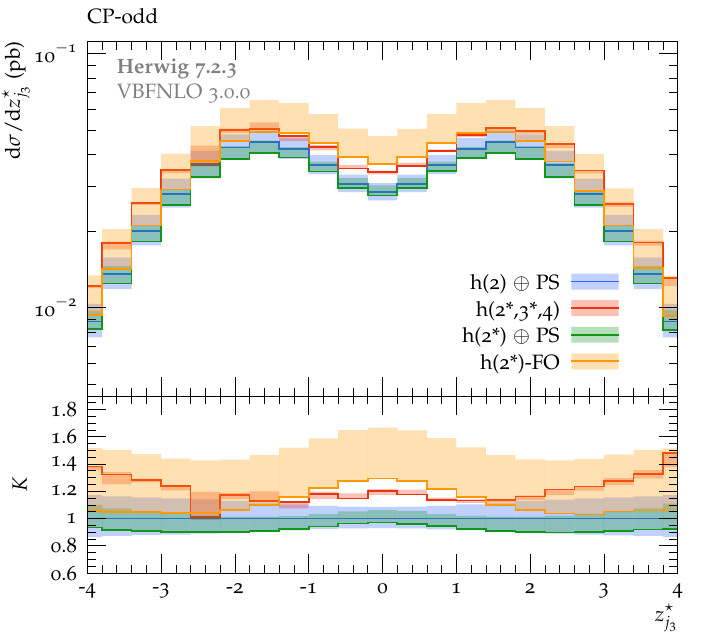}
  \end{subfigure}

  \caption{ The distribution of the transverse momentum of the third jet $p_{T,j_3}$ (left) and third-jet centrality variable, $z^{\star}_{j_3}$ (right), for the CP-mixed (top), CP-even (middle), and CP-odd (bottom) scenarios under INCL selection cuts. The distribution compares the matched predictions \hjjlops and \hjjnlops with the merged \hjjmerge prediction and the fixed-order \hjjnlo calculation. The corresponding ratios $K$ compared to the \hjjlops prediction are shown in the lower panels, with the shaded bands indicating the corresponding theoretical uncertainties.}
  \label{fig:B_j3_error}
\end{figure*}

The left-hand side of Fig.~\ref{fig:B_j3_error} shows the transverse momentum distribution of the third jet, $p_{T,j_3}$, for the CP-even, CP-mixed, and CP-odd scenarios. In all three cases, the distribution falls with increasing transverse momentum and exhibits a similar overall shape. The merged prediction \hjjmerge is larger than both the LO and NLO matched setups over the entire kinematic range, with deviations reaching approximately $30\%-40\%$ when $p_{T,j_3}>100 \text{GeV}$. The fixed-order calculation \hjjnlo shows the largest theoretical error band, which covers the merged results and differ from the matched results. The theoretical errors of \hjjnlops are largely covered by the uncertainty band of the LO matched setup \hjjlops. The shape of the distribution is similar for the CP-even, CP-mixed, and CP-odd scenarios, indicating that the third-jet transverse momentum is primarily sensitive to the modeling of additional QCD radiation rather than to the CP structure of the $HVV$ interaction. Consequently, while $p_{T,j_3}$ provides a useful observable for assessing the performance of merging and matching prescriptions, it offers limited discriminating power between different CP hypotheses.
Plots in the right-hand side of Fig.~\ref{fig:B_j3_error} present the normalized distribution of the third-jet centrality variable, $z^\star_{j_3}$ for the CP-mixed, CP-even, and CP-odd scenarios, where $z^\star_{j_3}$ is defined similarly to the so-called
Zeppenfeld variable \cite{Chen:2021phj,Jager:2020hkz,Rainwater:1996ud},
\begin{equation}
    z^{\star}_{j_3}=\frac{y_{j_3}-\frac{y_{j_1}+y_{j_2}}{2}}{\Delta y_{j_1j_2}}.
\end{equation}
The distributions are approximately symmetric about $z^{\star}_{j_3}=0$ and exhibit a suppression in the central region in general for all setups. The overall shape is similar for all three coupling configurations, indicating only a weak dependence of this observable on the CP structure of the anomalous $HVV$ interaction under the INCL selection cuts. The two matched predictions, \hjjlops and \hjjnlops, display comparable shapes throughout the region, although the latter yields a systematically smaller theoretical uncertainty. The merged \hjjmerge prediction lies above the matched results over most of the distribution, reflecting the impact of higher-multiplicity matrix elements on the description of additional jet radiation. The fixed-order \hjjnlo gives larger uncertainties, particularly in the central region compared to other setups among all CP configurations. By contrast, merged calculations exhibit reduced uncertainties and a more stable description across the full $z^{\star}_{j_3}$ range. The localized fluctuations in individual bins, especially for the CP-mixed and CP-odd configurations, are due to the statistical precision of the event samples.

\subsection {Impact of the CP structure on kinematic distributions}

\begin{figure*}[!tp]
  \centering

  \begin{subfigure}[t]{0.48\textwidth}
    \includegraphics[width=0.9\textwidth]{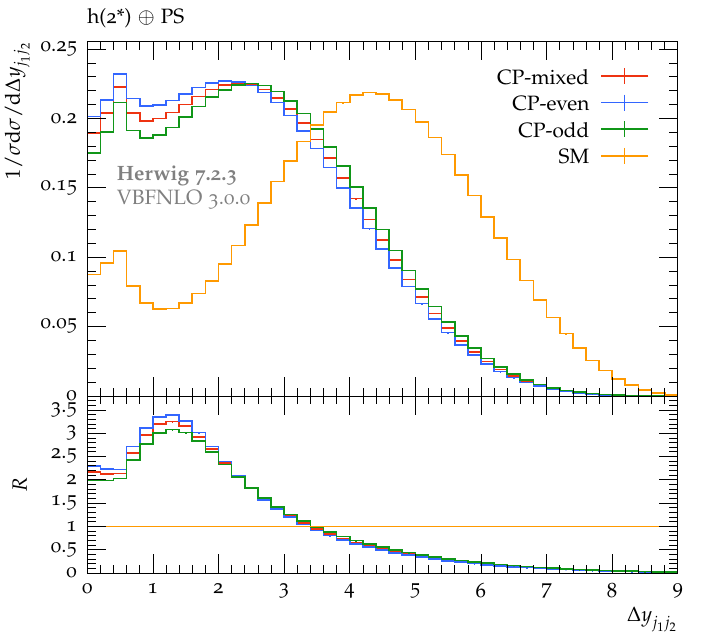}
  \end{subfigure}
  \hfill
  \begin{subfigure}[t]{0.48\textwidth}
    \includegraphics[width=0.9\textwidth]{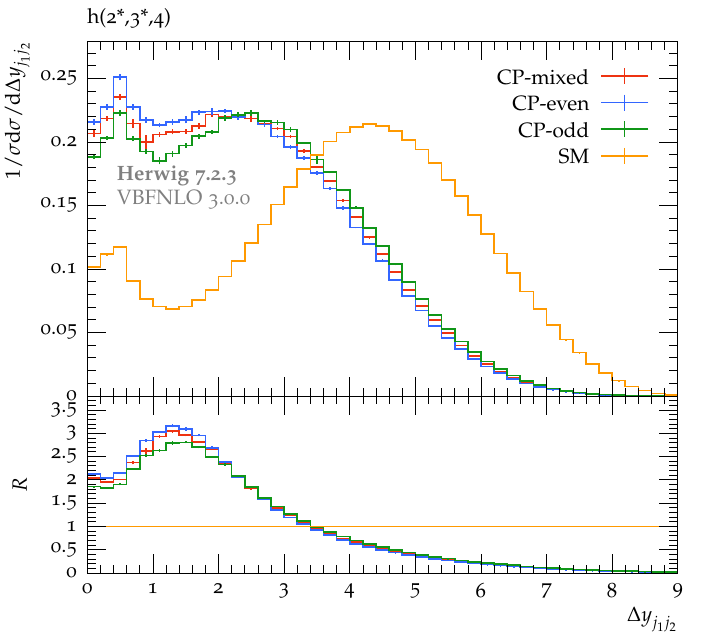}
  \end{subfigure}


  \begin{subfigure}[t]{0.48\textwidth}
    \includegraphics[width=0.9\textwidth]{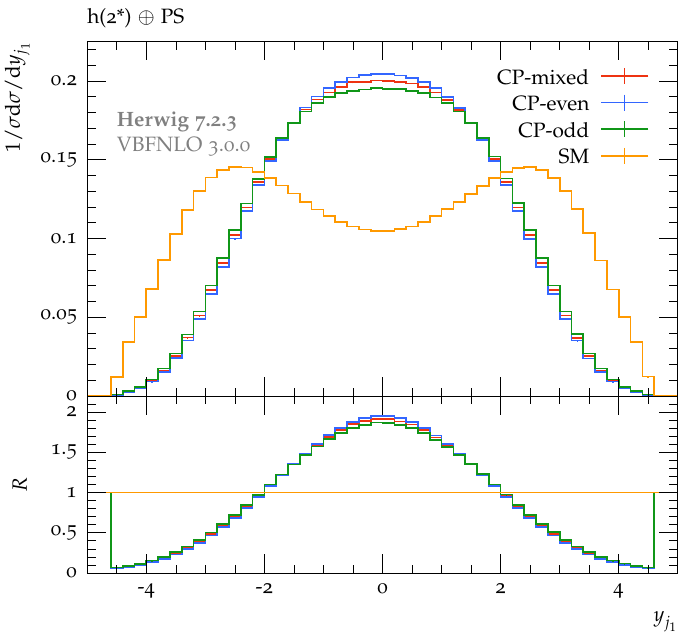}
  \end{subfigure}
  \hfill
  \begin{subfigure}[t]{0.48\textwidth}
    \includegraphics[width=0.9\textwidth]{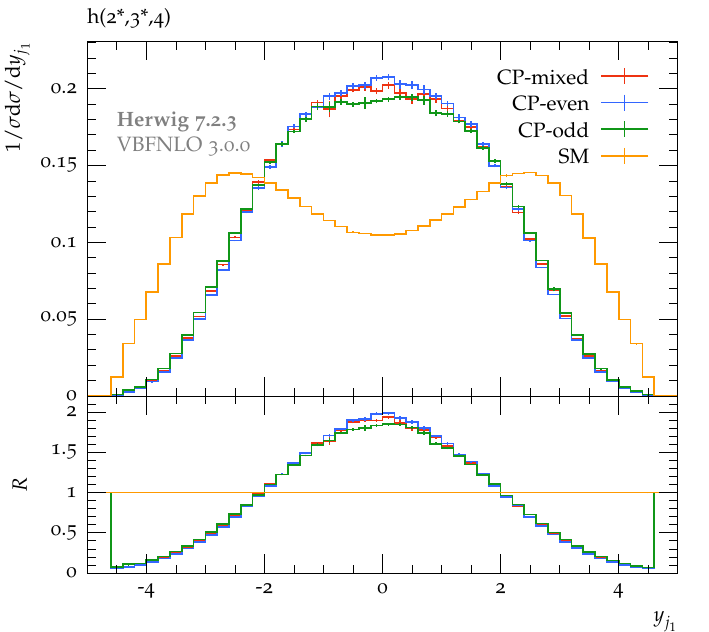}
  \end{subfigure}


  \begin{subfigure}[t]{0.48\textwidth}
    \includegraphics[width=0.9\textwidth]{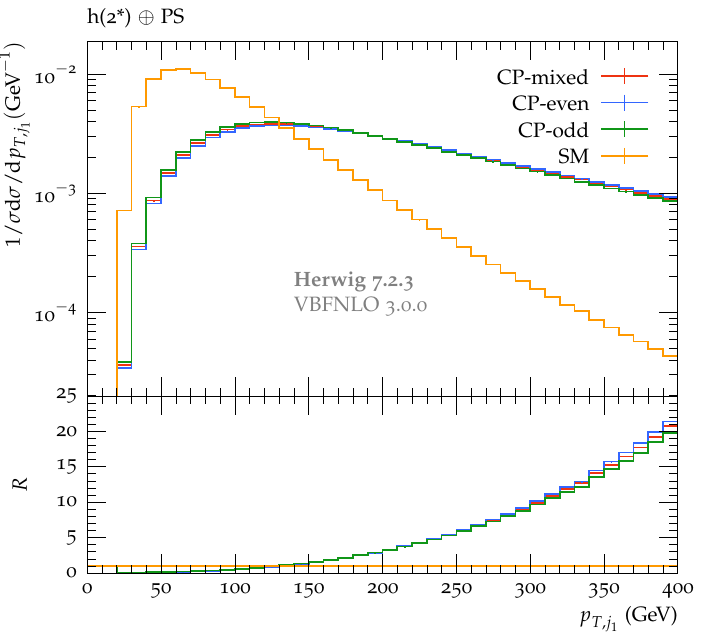}
  \end{subfigure}
  \hfill
  \begin{subfigure}[t]{0.48\textwidth}
    \includegraphics[width=0.9\textwidth]{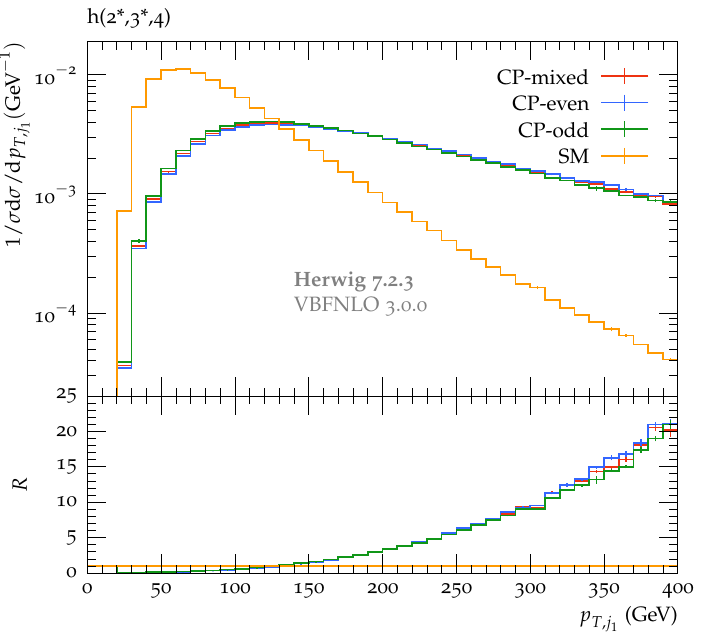}
  \end{subfigure}

  \caption{Normailized Distribution of rapidity separation between two leading jet $\Delta y_{j_1j_2}$ (top), rapidity of the leading jet $y_{j_1}$ (middle), and transverse momentum of the leading jet $p_{T,{j_1}}$ (bottom) for \hjjnlops (left) and \hjjmerge (right) setup with INCL cuts. Predictions for the anomalous coupling scenarios CP-mixed, CP-even, and CP-odd and SM are compared. The lower panels show the ratio $R$ with respect to the SM prediction.}
  \label{fig:C_jet1}
\end{figure*}

This section presents a detailed comparison of the three anomalous coupling scenarios, CP-even, CP-odd, and CP-mixed, with the SM predictions with INCL cuts. The normalized distribution plots focus on specific kinematic observables: the rapidity gap between the two leading jets $\Delta y_{j_1j_2}$, the rapidity distribution of the leading jet $y_{j_1}$, and the transverse momentum distribution $p_{T,j_1}$ of the leading jet. The ratio factor $R$ is defined in Eq.~\ref{eq:r}. 
The top figures in Fig.~\ref{fig:C_jet1} illustrate the distributions of $\Delta y_{j_1j_2}$ for the matched setup \hjjnlops and the merged setup \hjjmerge, respectively. All anomalous coupling scenarios show notable deviations from the SM predictions for the full range in $\Delta y_{j_1j_2}$. The SM distribution is shifted toward larger values of $\Delta y_{j_1j_2}$, while all the CP predictions populate a peak around $\Delta y_{j_1j_2}=1.4$ and fall more rapidly than SM predictions after $\Delta y_{j_1j_2}=3.4$. 
The normalized rapidity distribution of the leading jet, $y_{j_1}$, is shown in the middle panels of Fig.~\ref{fig:C_jet1}. $y_{j_1}$ distributions associated with the anomalous couplings appear to be more centrally distributed and peak around $y_{j_1}=0$, whereas the SM exhibits a broader rapidity distribution. 
The bottom of Fig.~\ref{fig:C_jet1} presents the normalized transverse momentum distribution of the first leading jet, $p_{T,j_1}$. The normalized distribution shows separation from the beginning, and the ratio with SM increases steadily after $p_{t,j_1}>150$~GeV. These features are present both in the \hjjnlops and in the \hjjmerge predictions. However, despite observable differences from the SM, the transverse momentum distribution alone does not distinctly separate the different CP scenarios. The distributions in Fig.~\ref{fig:C_jet1} show clear differences between the SM prediction and the CP-modified interactions, while the CP-even, CP-odd, and CP-mixed scenarios remain very similar to each other.

\subsection{Comparison of Form Factor Choices}

\begin{figure*}[!tp]
  \centering
  \includegraphics[width=0.432\textwidth]{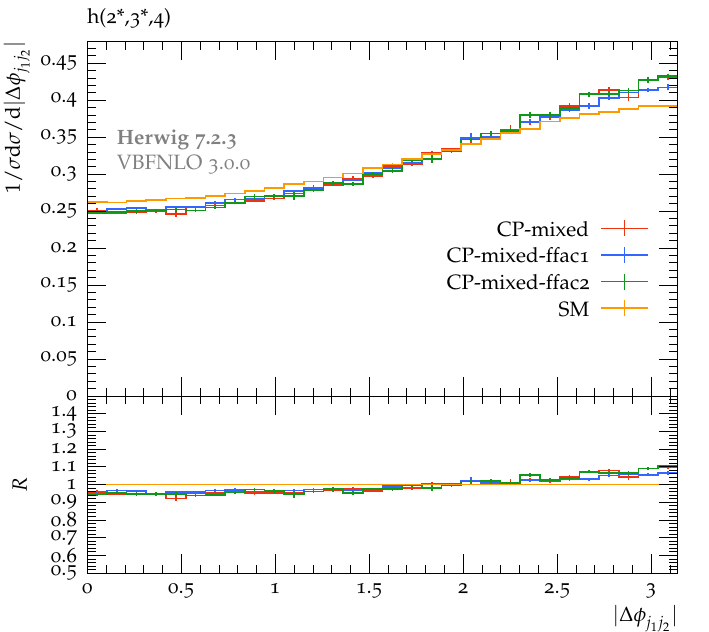}
  
  \includegraphics[width=0.432\textwidth]{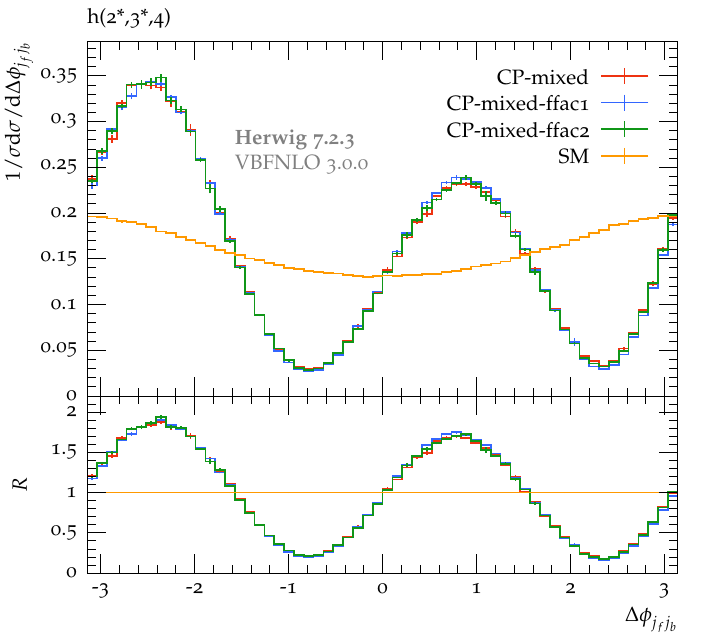}
  \includegraphics[width=0.432\textwidth]{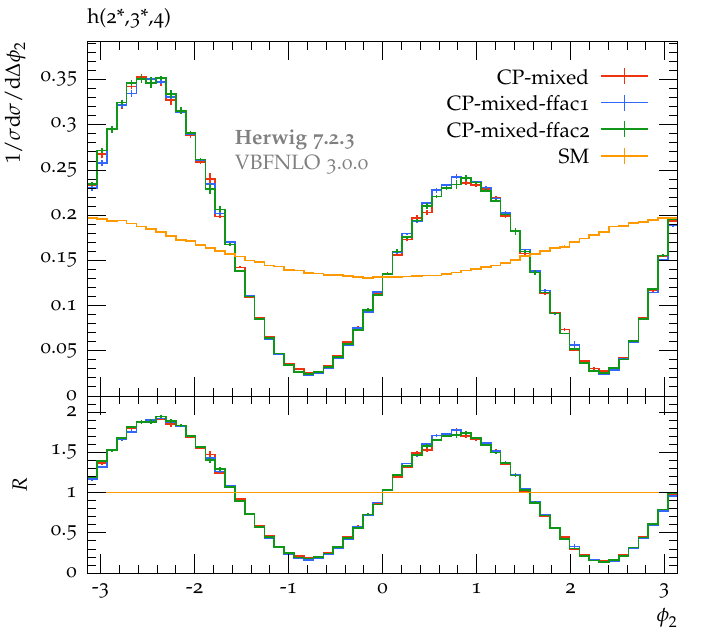}
  
  \caption{Normalized distribution of $|\Delta\phi_{j_1j_2}|$ (top), $\Delta\phi_{j_fj_b}$ (bottom left), and $\Delta\phi_2$ (bottom right) using merging \hjjmerge setup with TIGHT cuts. The normalized distribution compared against the SM is the anomalous coupling with the CP-mixed case, the form factor with parametrization of $F_1$ in Eq.\ref{eq:ffac}, and the form factor with parametrization of $F_2$ in Eq.\ref{eq:ffac}, where the characteristic mass scale is set to $\Lambda=400 \GeV$.}
  \label{fig:D_phi2}
\end{figure*}

We also investigate the effect of the form factor (FF) for anomalous couplings that are parametrized through a form factor that multiplies the CP-mixed coupling \cite{Figy:2004pt, Hankele:2006ma}, in the form of 
\begin{align}
    F_1&=\frac{\Lambda^2}{q_1^2-\Lambda^2}\frac{\Lambda^2}{q_2^2-\Lambda^2} \nonumber\\
    F_2&=-2\Lambda^2C_0(q_1^2,q_2^2,(q_1+q_2)^2, \Lambda^2)
    \label{eq:ffac}
\end{align}
where $\Lambda$ corresponds to the mass scale of new physics and $C_0$ is the usual scalar loop integral for triangle graphs. The impact of these form factor choices is studied for the normalized distributions using the \hjjmerge prediction. Here we set the mass scale $\Lambda=400~\text{GeV}$ with the different choices denoted by CP-mixed-ffac1 using $F_1$ and \text{CP-mixed-ffac2} using $F_2$ in Eq.\ref{eq:ffac}. Using the alternative form-factor prescription, the fiducial cross section is found to be $\sigma_{F_1} = 7.990229\times 10^{-2} \pm 1.033484\times 10^{-4}~\mathrm{pb}$ and $\sigma_{F_2} = 1.446580\times 10^{-1} \pm 2.706945\times 10^{-4}~\mathrm{pb}$ after the TIGHT selection cuts.
The signed observables \dphijfjb and $\phi_2$ in Fig.~\ref{fig:D_phi2} show the characteristic CP-sensitive shape, which the CP-mixed result exhibits pronounced maxima and minima across the full range. The two form-factor variations closely follow the nominal CP-mixed curve, indicating that the observable are insensitive to the tested form-factor scale. This is also reflected in the ratio plot. For \dphijfjb and $\phi_2$, all ratios displays large oscillations relative to the SM, but the CP-mixed, CP-mixed-ffac1, and CP-mixed-ffac2 predictions remain almost indistinguishable. The unsigned observable \dphijj shows a similar trend with the small differences due to statistical uncertainty. 

\begin{figure*}[!tp]
  \centering
  \begin{subfigure}[t]{0.48\textwidth}
  \includegraphics[width=0.9\textwidth]{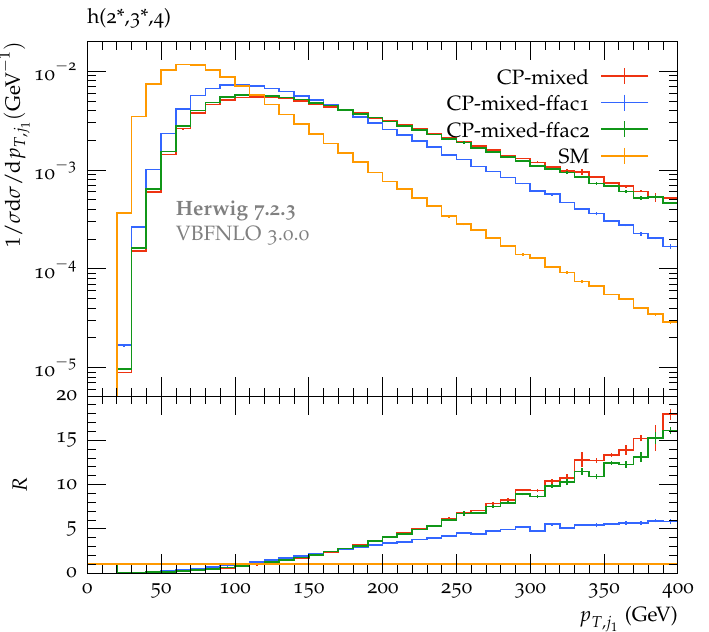}
  \end{subfigure}
  \vspace{0.5em}
  \begin{subfigure}[t]{0.48\textwidth}
  \includegraphics[width=0.9\textwidth]{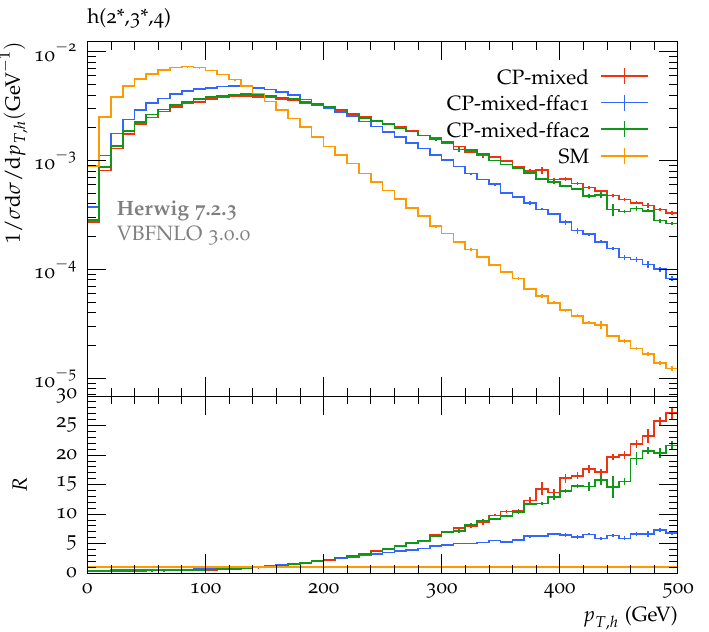}
  \end{subfigure}
  \caption{Normalized distribution of transverse momentum of the hardest jet $p_{T,j_1}$ (left) and the transverse momentum of Higgs boson $p_{T,h}$ (right) using \hjjmerge setup with TIGHT cuts. The distributions compared against the SM are the anomalous coupling with CP-mixed case, the form factor with parametrization of $F_1$ and $F_2$ in Eq.\ref{eq:ffac}, where the characteristic mass scale is set to $\Lambda=400 \GeV$.}
  \label{fig:D_jet1}
\end{figure*}

Fig.~\ref{fig:D_jet1} shows the normalized transverse momentum distributions for the hardest jet $p_{T,j_1}$ (left) and the Higgs boson $p_{T,h}$ (right) for the \hjjmerge prediction. Compared with the SM result, the CP-mixed interaction produces different shapes in both the Higgs transverse momentum $p_{T,h}$ and the leading-jet transverse momentum $p_{T,j_1}$. The difference between CP-mixed and two FF cases is small in the low-$p_T$ region $p_{T,j_1}<200~\text{GeV}$ and $p_{T,h}<250~\text{GeV}$, where all predictions have similar shapes, but becomes increasingly apparent in the high-$p_T$ tail. In both plots, the CP-mixed prediction rises rapidly above the SM baseline. In the high-$p_T$ region, CP-mixed-ffac1 shows a stronger suppression compared to the CP-mixed results, while CP-mixed-ffac2 remains closer to the CP-mixed prediction. The CP-mixed-ffac2 prediction exhibits a noticeable deviation in the normalized transverse momentum distributions for $p_{T,j_1}>350~\text{GeV}$ and $p_{T,h}>400~\text{GeV}$. These results show that, unlike the normalized azimuthal-angle distributions, both observables $p_{T,j_1}$ and $p_{T,h}$ are more sensitive to the form factor effect.

\section{\label{sec:conc} Conclusions}

In this work, we have presented a detailed study of Higgs boson production via vector boson fusion in the presence of anomalous $HVV$ couplings, employing state-of-the-art NLO QCD matrix elements from \vbfnlo consistently matched and merged with parton showers in \herwig. Our analysis has focused on the interplay between anomalous CP structures and higher-order QCD radiation, with particular emphasis on jet-related and CP-sensitive observables.

We have demonstrated that azimuthal-angle observables, including $|\Delta\phi_{j_1j_2}|$, $\Delta\phi_{j_fj_b}$, and $\phi_2$, provide robust discrimination between CP-even, CP-odd, and CP-mixed Higgs interactions. These observables are remarkably stable under the inclusion of additional QCD radiation, with differences between matched, merged, and fixed-order predictions remaining within theoretical uncertainties at the level of $\mathcal{O}(20\%)$. This confirms their reliability for probing the CP structure of the $HVV$ vertex.

In contrast, observables sensitive to additional jet activity, such as $z^\star_{j_3}$, show a strong dependence on the treatment of higher-order radiation. In particular, we find that multi-jet merging is essential for accurately describing regions with significant additional jet activity, where fixed-order predictions fail, and matched simulations exhibit sizeable deviations. This highlights the importance of merged simulations for realistic collider analyses.

When comparing anomalous coupling scenarios to the Standard Model, we observe clear deviations in rapidity and transverse-momentum distributions, although these observables do not individually distinguish between different CP hypotheses. Furthermore, we have investigated the impact of form factors and found that while azimuthal observables remain largely insensitive to their presence, high-$p_T$ distributions exhibit noticeable suppression effects, particularly in the tails.

Overall, our study demonstrates that NLO multi-jet merging provides stable and reliable predictions for VBF Higgs production in the presence of anomalous couplings, and is crucial for accurately modeling jet radiation effects. The robustness of CP-sensitive observables against higher-order corrections makes them powerful tools for experimental analyses. Our results provide important guidance for future precision measurements of the Higgs sector at the LHC and for ongoing searches for CP-violating effects.

Future work may include the incorporation of hadronization and MPI effects, a systematic study of theoretical uncertainties associated with merging scales, and the extension to a full SMEFT framework including additional operators.

\begin{acknowledgments}
We would like to thank BeoShock HPC Center at Wichita State University, Beocat HPC Center at Kansas State University, and the University of Delaware IT Research Cyberinfrastructure for all the numerical computations.

We would also like to thank Eric Orona for his contributions to this project. Erick was funded through the Applied Learning in Physics Research Program at Wichita State University. 

A special thanks goes to Micheal Rauch for making an improvement to the Higgs boson plus three jet process in VBFNLO. These improvements were instrumental in ensuring the success of this project. Lastly, we would like to thank Simon Pl\"{a}tzer for reading through the draft and for setting up the {\tt Matchbox} framework inside {\tt Herwig 7}. Without this type of software development effort, calculations such as the one presented in this article would not be possible.

\end{acknowledgments}

\newpage
\bibliographystyle{apsrev4-2}
\bibliography{refern}

\end{document}